\documentclass[aps,prd,superscriptaddress,nofootinbib,tighten,preprint]{revtex4}
\pdfoutput=1
\usepackage[utf8]{inputenc}
\usepackage[english]{babel}
\usepackage{amsmath}
\usepackage{subfigure}
\usepackage{amsfonts}
\usepackage{amssymb}
\usepackage{graphicx}
\newcommand{\be}{\begin{equation}}
\newcommand{\ee}{\end{equation}}
\newcommand{\bea}{\begin{eqnarray}}
\newcommand{\eea}{\end{eqnarray}}

\newcommand{\umt}{{\rm U(1)}_{L_{\mu}-L_{\tau}}}
\newcommand{\vmt}{v_{\mu \tau}}
\newcommand{\gmt}{g_{\mu \tau}}
\newcommand{\zmt}{Z_{\mu \tau}}
\newcommand{\mzmt}{M_{\zmt}}
\newcommand{\dm}{\phi_{DM}}
\newcommand{\dmd}{\phi^\dagger_{DM}}

\newcommand{\mdm}{M_{DM}}
\newcommand{\ldh}{\lambda_{Dh}}
\newcommand{\ldH}{\lambda_{DH}}
\newcommand{\nmt}{n_{\mu \tau}}

\newcommand{\lm}{L_{\mu}}
\newcommand{\lt}{L_{\tau}}
\newcommand{\nre}{N_e}
\newcommand{\nrm}{N_{\mu}}
\newcommand{\nrt}{N_{\tau}}
\newcommand{\smgauge}{{\rm SU}(2)_{\rm L}\times {\rm U}(1)_{\rm Y}}
\newcommand{\nn}{\nonumber}

\catcode`@=12
\def\la{\mathrel{\mathchoice {\vcenter{\offinterlineskip\halign{\hfil
$\displaystyle##$\hfil\cr<\cr\sim\cr}}}
{\vcenter{\offinterlineskip\halign{\hfil$\textstyle##$\hfil\cr<\cr\sim\cr}}}
{\vcenter{\offinterlineskip\halign{\hfil$\scriptstyle##$\hfil\cr<\cr\sim\cr}}}
{\vcenter{\offinterlineskip\halign{\hfil$\scriptscriptstyle##$\hfil\cr<\cr\sim
\cr}}}}}
\def\ga{\mathrel{\mathchoice {\vcenter{\offinterlineskip\halign{\hfil
$\displaystyle##$\hfil\cr>\cr\sim\cr}}}
{\vcenter{\offinterlineskip\halign{\hfil$\textstyle##$\hfil\cr>\cr\sim\cr}}}
{\vcenter{\offinterlineskip\halign{\hfil$\scriptstyle##$\hfil\cr>\cr\sim\cr}}}
{\vcenter{\offinterlineskip\halign{\hfil$\scriptscriptstyle##$\hfil\cr>\cr\sim
\cr}}}}}
\begin{document}

\title{Neutrino Mass, Dark Matter and Anomalous Magnetic Moment
of Muon in a $\umt$ Model}

\author{Anirban Biswas}
\email{anirbanbiswas@hri.res.in}
\affiliation{Harish-Chandra Research Institute, Chhatnag Road,
Jhunsi, Allahabad 211 019, India}
\affiliation{Homi Bhabha National Institute, Training School Complex, Anushaktinagar, Mumbai - 400094, India}
\author{Sandhya Choubey}
\email{sandhya@hri.res.in}
\affiliation{Harish-Chandra Research Institute, Chhatnag Road,
Jhunsi, Allahabad 211 019, India}
\affiliation{Homi Bhabha National Institute, Training School Complex, Anushaktinagar, Mumbai - 400094, India}
\affiliation{Department of Theoretical Physics, School of
Engineering Sciences, KTH Royal Institute of Technology, AlbaNova
University Center, 106 91 Stockholm, Sweden}
\author{Sarif Khan}
\email{sarifkhan@hri.res.in}
\affiliation{Harish-Chandra Research Institute, Chhatnag Road,
Jhunsi, Allahabad 211 019, India}
\affiliation{Homi Bhabha National Institute, Training School Complex, Anushaktinagar, Mumbai - 400094, India}

\begin{abstract} 

The observation of neutrino masses, mixing and the
existence of dark matter are amongst the most important
signatures of physics beyond the Standard Model (SM). 
In this paper, we propose to extend the SM by a local 
$L_\mu - L_\tau$ gauge symmetry, two additional complex
scalars and three right-handed neutrinos. The $L_\mu - L_\tau$
gauge symmetry is broken spontaneously when one of the scalars
acquires a vacuum expectation value. The $L_\mu - L_\tau$
gauge symmetry is known to be anomaly free and can explain
the beyond SM measurement of the anomalous muon $({g-2})$
through additional contribution arising from the extra
$Z_{\mu\tau}$ mediated diagram. Small neutrino masses are
explained naturally through the Type-I seesaw mechanism, 
while the mixing angles are predicted to be in their
observed ranges due to the broken $L_\mu-L_\tau$ symmetry.
The second complex scalar is shown to be stable and
becomes the dark matter candidate in our model. 
We show that while the $Z_{\mu\tau}$ portal is ineffective
for the parameters needed to explain the anomalous
muon $({g-2})$ data, the correct dark matter relic
abundance can easily be obtained from annihilation through
the Higgs portal. Annihilation of the scalar 
dark matter in our model can also explain the Galactic Centre
gamma ray excess observed by Fermi-LAT. We show the predictions
of our model for future direct detection experiments and
neutrino oscillation experiments. 
\end{abstract}
\maketitle
\section{Introduction}
\label{Intro}
Explaining of the origin of nonzero neutrino masses and
dark matter (DM) are two of the principal challenges which
theoretical high energy physics has been facing
over the last few decades. 
Neutrinos were predicted to be massless in the Standard Model (SM)
of particle physics. However, in 1998 the neutrino
oscillation (oscillation between mass and flavour eigenstates)
which requires nonzero mass differences between
different generation of neutrinos and mixing between them, was 
unambiguously observed  by the Super-Kamiokande atmospheric
neutrino experiment \cite{Fukuda:1998mi}. Existence of neutrino
mass and mixing requires the extension of the SM. 
Neutrino oscillations have now been established at a
very high confidence level by many outstanding
experimental observations by experiments such as SNO
\cite{Ahmad:2002jz} (solar neutrino experiment), 
KamLand \cite{Eguchi:2002dm} (reactor neutrino experiment),
Daya Bay \cite{An:2015nua}, RENO \cite{RENO:2015ksa},
Double Chooz \cite{Abe:2014bwa} (reactor neutrino experiments
with short baselines), 
T2K \cite{Abe:2015awa, Salzgeber:2015gua} and NO$\nu$A
\cite{Adamson:2016tbq, Adamson:2016xxw} (accelerator neutrino experiments). 
At present for normal (inverted) mass ordering 
scenarios, the best fit values \cite{Capozzi:2016rtj} of 
neutrino oscillation parameters obtained
from global neutrino oscillation data are: 
\footnote{We 
define $\Delta m^2_{ij}=m_i^2-m_j^2$. The mass squared difference 
$\Delta m^2_{{\rm atm}} = m_{3}^2 - ((m_2^{2}+m_1^{2})/2)$, 
where we use the notation given in \cite{Capozzi:2016rtj}.}
\begin{eqnarray}
&\Delta m^2_{21} = 7.37\times 10^{-5} ~{\rm eV}^2,~~
|\Delta m^2_{{\rm atm}}| = 2.50\,(2.46) \times 10^{-5}~ {\rm eV}^2&\nonumber \\
&\theta_{12} = 33.02^\circ,~
\theta_{23} = 41.38^\circ\,(48.97^\circ),~
\theta_{13} = 8.41^\circ\,(8.49^\circ)
\label{eq:mix}
\end{eqnarray}

On other hand, the existence of dark matter
in the Universe has been confirmed to a very
high statistical significance by many indirect
evidences such as the flatness of rotation curves of spiral
galaxies \cite{Sofue:2000jx}, collision of galaxies in a galaxy cluster
(bullet cluster and others) \cite{Clowe:2003tk, Harvey:2015hha}, 
gravitational lensing \cite{Bartelmann:1999yn} and the measurements of the 
Cosmic Microwave Background (CMB) \cite{Hinshaw:2012aka,Ade:2015xua}. 
The satellite borne CMB experiments, 
WMAP \cite{Hinshaw:2012aka} and Planck \cite{Ade:2015xua},
have measured the fractional contribution of
dark matter to the present energy density of the Universe
(commonly known as DM relic density) to be around  
0.25 with an extremely good accuracy, while the contribution of the 
visible baryonic matter is only around 0.05.
The rest $\sim 70$\% of energy density of the
Universe is also coming from an mysterious energy
called the Dark Energy \cite{Copeland:2006wr}. The current best observed
value of DM relic density is \cite{Ade:2015xua}
\begin{equation}
\Omega_{\rm DM} h^2= 0.1197\pm0.0022
\,.
\end{equation}

Like the neutrino sector mentioned before,
the SM of particle physics does not have any
stable particle(s) which can play the role
of viable DM candidate(s). Therefore beyond
Standard Model (BSM) scenario is required
to explain these two long standing puzzles.
Weakly Interacting Massive Particles (WIMP)
\cite{Gondolo:1990dk, Srednicki:1988ce} have been proposed as 
one of the most promising candidates to explain the dark matter 
puzzle of the Universe. Many direct detection experiments like
LUX \cite{Akerib:2015rjg}, XENON \cite{Aprile:2015uzo}
and CDMS \cite{Agnese:2014aze} have been
trying to detect WIMPs through their spin independent
as well as spin dependent elastic scattering with the
detector nuclei. However, no convincing signature of WIMPs
has been observed yet in the direct detection experiments, 
giving bounds on the WIMP-nucleon scattering cross section. 
Recently, the LUX collaboration has reported the most stringent
upper bound on DM-nucleon spin independent scattering cross section
to be around $2.2\times10^{-46}$ cm$^{-2}$ \cite{lux2016}
for a $\sim 50$ GeV DM particle. 

Signature of DM can also appear in indirect detection experiments,
looking for high energy neutrinos, gamma rays and charged cosmic rays
(electrons, positrons, protons and antiprotons) coming from the annihilation
or decay of DM particles \cite{Hooper:2009zm}. 
In this work, we will briefly discuss about the Galactic Centre 
gamma-ray excess in the energy range 1-3 GeV which has been
observed by the Fermi-LAT collaboration \cite{Atwood:2009ez}.
Although, there are some astrophysical explanations
such as unresolved point sources (e.g. millisecond pulsar)
\cite{Lee:2015fea, Bartels:2015aea} for this excess
gamma-ray flux, but in this work we will explain this
anomalous excess by the process of DM annihilation into $b\bar{b}$ final state.
The authors of Ref. \cite{Calore:2014nla} have given
constraints on DM mass and its annihilation cross section
${\langle \sigma {\rm v}_{b\bar{b}} \rangle}$ to explain
the gamma-ray excess which are  $48.7^{+6.4}_{-5.2}$ GeV and
$1.75^{+0.28}_{-0.26}\times 10^{-26}$ cm$^3/$s for the $b\bar{b}$
annihilation channel respectively. In the present model
we can explain this excess gamma-ray flux in the energy range 1-3 GeV.

The SM has accidental U(1) global symmetries like
the baryon ($B$) and the lepton number ($L$) conservation. However,
if we want to convert these global symmetries into a
local one then they become anomalous. The anomaly free
situation can be obtained if instead of considering
$B$ and $L$ separately one uses some combinations
between them. There are only four non-anomalous combinations 
possible, and these are ${B-L}$, ${L_{e}-\lm}$,
${\lm-\lt}$ and ${L_{e}-\lt}$ where
${L}_{e}$, $\lm$ and $\lt$ are the respective
lepton numbers of generations associated with leptons
$e$, $\mu$ and $\tau$ while $L=L_{e}+\lm+\lt$ is the
total lepton number. Out of these four possible
combinations, axial vector anomaly
\cite{Adler:1969gk, Bardeen:1969md} and
gravitational gauge anomaly
\cite{Delbourgo:1972xb, Eguchi:1976db} of local
$B-L$ symmetric models can be cancelled by
the introduction of extra chiral fermions to the 
SM such as three right handed neutrinos \cite{Okada:2010wd} or
two left and right handed singlet fermions with
appropriate $B-L$ charges \cite{Patra:2016ofq}. However, unlike
the $B-L$ case, the anomaly cancellation does
not require any extra chiral fermionic degrees of freedom for
the last three cases where the linear combinations of
different generational lepton numbers \cite{He:1990pn,
He:1991qd, Ma:2001md} are considered.
Here anomalies cancel between different leptonic generations.
Among these three possible scenarios U(1)$_{\lm-\lt}$
extension \cite{Xing:2015fdg, Brignole:2004ah, Mohapatra:2005yu, Xing:2006xa,
Adhikary:2006rf, Kitabayashi:2005fc, Baek:2008nz, He:2011kn, Ge:2010js,
EstebanPretel:2007yq, Heeck:2011wj, Joshipura:2009tg, Fuki:2006ag, Aizawa:2005yy,
Adhikary:2009kz, Grimus:2012hu, Grimus:2005jk, Rodejohann:2005ru,
Haba:2006hc, Joshipura:2009fu, Xing:2010ez,
Bandyopadhyay:2009xa, Grimus:2006jz, Fukuyama:1997ky,
Araki:2010kq, Feng:2012jn, Kile:2014jea, Kim:2015fpa,
Zhao:2016orh, Patra:2016shz}
of SM is less constrained as in this case the extra neutral gauge boson
does not couple to electron and quarks and
therefore $\zmt$ is free from any constraints
coming from lepton and hadron colliders such
as LEP \cite{Carena:2004xs, Cacciapaglia:2006pk}
and LHC \cite{Aad:2014cka}. Therefore, the mass of
$\zmt$ can be as light as $\mathcal{O}$ (100 MeV) for a low
value of gauge coupling $\gmt \la 10^{-3}$ which is
required to satisfy the constraints arising
from neutrino trident production \cite{Altmannshofer:2014pba}. 
One of the phenomenological motivation for the U(1)$_{\lm-\lt}$
extension of the SM is that it can explain the muon ($g-2$)
anomaly between the theoretical value predicted by
the SM \cite{Agashe:2014kda} which is $a_{\mu}^{\rm th} =
1.1659179090(65)\times 10^{-3}$ and the experimental value
\cite{Jegerlehner:2009ry} which is $a_{\mu}^{\rm exp} = 1.16592080(63)
\times 10^{-3}$. The difference between theoretical
and experimantal value \cite{Jegerlehner:2009ry} is,
\begin{eqnarray}
\Delta a_{\mu} = a_{\mu}^{\rm exp} - a_{\mu}^{\rm th}
= (29.0 \pm 9.0) \times 10^{-10}\,.
\label{delta}
\end{eqnarray}

In this work, we have considered the gauged
U(1)$_{\lm-\lt}$ extension of the SM. Amongst the main motivations 
for our choice of this model is that it provides $\mu-\tau$ flavor 
symmetry which could naturally explain the peculiar neutrino mixing parameters 
(cf. Eq.\,\,(\ref{eq:mix})) wherein $\theta_{23}$ is close to maximal and 
$\theta_{13}$ is small. As mentioned above, this model can also 
explain the muon ($g-2$) anomaly \cite{Baek:2001kca, Bennett:2004pv,
Mohanty:2013soa, Das:2014kwa, Altmannshofer:2016brv} 
for a range of $\zmt$ mass and $\gmt$ 
consistent with collider constraints. We will further extend
this model with a complex scalar, which will become a viable DM
candidate. $\umt$ extended
Ma model \cite{Ma:2006km} has been studied earlier
in the context of small neutrino mass generation in one
loop level \cite{Baek:2015mna} and dark matter \cite{Baek:2015fea}.
A review on earlier works about $\mu-\tau$ flavour symmetry in
neutrino sector can be found in \cite{Xing:2015fdg} and references therein.
In order to generate neutrino masses through the Type-I seesaw mechanism
\cite{Minkowski:1977sc, Yanagida:1979as, Mohapatra:1979ia, Schechter:1980gr}
in the present scenario, we have introduced three right handed
neutrinos ($N_e$, $N_{\mu}$, $N_{\tau}$) with $\lm-\lt$ charges
0, 1 and -1 respectively in the fermionic sector of SM. The scalar sector of the model is
also enlarged by the addition of two complex scalar singlets
($\phi_H$ and $\dm$) with nonzero $\lm-\lt$ charge. The proposed
$\lm-\lt$ symmetry is broken spontaneously when $\phi_H$ acquires
vacuum expectation value (VEV) $\vmt$ and thereby making $\zmt$ massive. 
The breaking of $\lm-\lt$ symmetry also results in
additional terms in the neutrino mass matrix. In particular,
the $\mu-\tau$ symmetry is broken and we can generate neutrino
masses and mixing parameters consistent with current bounds.
We show that the complex scalar $\phi_{DM}$ is stable in
our model and hence becomes the DM candidate satisfying 
the constraints from Planck, LUX and LHC results. We
show that a sub-region of the parameter space that is
consistent with Planck, LUX and LHC results can also
explain the Galactic Centre gamma ray excess observed by Fermi-LAT.

The rest of the article is organised as follows. In Section \ref{model},
we describe the model for the present work. In Section \ref{muong-2}
and Section \ref{numass} we discuss muon (${g-2}$) and neutrino
masses and mixing angles, respectively. In Section \ref{dm}
we study the DM constraints and its related phenomenology.
In section \ref{sandc} we conclude. 
\section{Model}
\label{model}

In this present work, we have considered a minimal extension of the SM 
where we have imposed an extra local $\umt$ symmetry to the
SM Lagrangian, where $\lm$ and $\lt$ denote 
the muon lepton number and tau lepton number respectively. Therefore, the
Lagrangian of the present model remains invariant under the
${\rm SU}(3)_{\rm c}\times \smgauge \times \umt$ gauge symmetry.
This model is free from axial vector and mixed gravitational
gauge anomalies as these anomalies cancel between second and
third generations of leptons without the requirement of
any additional chiral fermion. 
The full particle content of our model  
and their respective charges under $\smgauge\times\umt$ gauge groups
are listed in Tables \ref{tab1} and \ref{tab2}.
In order to break
the $\umt$ symmetry spontaneously, we need a complex
scalar field $\phi_H$ with a non-trivial $\lm-\lt$ charge assignment such that 
the $\lm-\lt$ symmetry is broken spontaneously when 
$\phi_H$ picks up a vacuum expectation value $\vmt$. 
Spontaneous breaking
of the $\lm-\lt$ symmetry generates mass for the extra neutral gauge boson $\zmt$. 
It has been shown that the spontaneously broken $\lm-\lt$ model 
can explain the anomalous muon $g-2$ signal.  
The $\lm-\lt$ symmetry is a flavor symmetry and hence can be used to 
explain the peculiar mixing pattern of the neutrinos \cite{Choubey:2004hn}. 
In our model we generate small neutrino masses through the
Type-I seesaw mechanism. To that end we introduce three
right handed neutrinos ($\nre$, $\nrm$, $\nrt$) with 
$\lm-\lt$ charges of 0, 1 and $-1$ respectively, such that their
presence do not introduce any further anomaly. In the $\umt$  symmetric
limit the right-handed neutrino mass has exact $\mu-\tau$ symmetry. We will
show that the spontaneous breaking of the gauged $\umt$ symmetry leads to additional 
terms in the right-handed neutrino mass matrix, providing a natural 
explanation of the neutrino masses and mixing parameters observed in 
neutrino oscillation experiments, given in Eq. (\ref{eq:mix}). 
We also add another complex scalar field $\dm$ in the model, 
with a chosen $\lm-\lt$ charge $\nmt$ such that the Lagrangian
does not contain any term with odd power of $\dm$. Also the scalar
field $\dm$ does not acquire any VEV and consequently in this model
$\dm$ becomes odd under a remnant $\mathbb{Z}_2$ symmetry
after the spontaneous breaking of the gauged $\umt$ symmetry, which
ensure its stability. Hence $\dm$ can be a viable dark matter
candidate. 
\begin{center}
\begin{table}[h!]
\begin{tabular}{||c|c|c|c||}
\hline
\hline
\begin{tabular}{c}
    Gauge\\
    Group\\ 
    \hline
    
    ${\rm SU(2)}_{\rm L}$\\ 
    \hline
    ${\rm U(1)}_{\rm Y}$\\ 
\end{tabular}
&

\begin{tabular}{c|c|c}
    \multicolumn{3}{c}{Baryon Fields}\\ 
    \hline
    $Q_{L}^{i}=(u_{L}^{i},d_{L}^{i})^{T}$&$u_{R}^{i}$&$d_{R}^{i}$\\ 
    \hline
    $2$&$1$&$1$\\ 
    \hline
    $1/6$&$2/3$&$-1/3$\\ 
\end{tabular}
&
\begin{tabular}{c|c|c}
    \multicolumn{3}{c}{Lepton Fields}\\
    \hline
    $L_{L}^{i}=(\nu_{L}^{i},e_{L}^{i})^{T}$ & $e_{R}^{i}$ & $N_{R}^{i}$\\
    \hline
    $2$&$1$&$1$\\
    \hline
    $-1/2$&$-1$&$0$\\
\end{tabular}
&
\begin{tabular}{c|c|c}
    \multicolumn{3}{c}{Scalar Fields}\\
    \hline
    $\phi_{h}$&$\phi_{H}$&$\phi_{DM}$\\
    \hline
    $2$&$1$&$1$\\
    \hline
    $1/2$&$0$&$0$\\
\end{tabular}\\
\hline
\hline
\end{tabular}
\caption{Particle contents and their corresponding
charges under SM gauge group.}
\label{tab1}
\end{table}
\end{center}
\begin{center}
\begin{table}[h!]
\begin{tabular}{||c|c|c|c||}
\hline
\hline
\begin{tabular}{c}
    Gauge\\
    Group\\ 
    \hline
    $\umt$\\ 
    
\end{tabular}
&
\begin{tabular}{c}
    \multicolumn{1}{c}{Baryonic Fields}\\ 
    \hline
    $(Q^{i}_{L}, u^{i}_{R}, d^{i}_{R})$\\ 
    \hline
    $0$ \\ 
    
\end{tabular}
&
\begin{tabular}{c|c|c}
    \multicolumn{3}{c}{Lepton Fields}\\ 
    \hline
    $(L_{L}^{e}, e_{R}, N_{R}^{e})$ & $(L_{L}^{\mu}, \mu_{R},
    N_{R}^{\mu})$ & $(L_{L}^{\tau}, \tau_{R}, N_{R}^{\tau})$\\ 
    \hline
    $0$ & $1$ & $-1$\\ 
    
\end{tabular}
&
\begin{tabular}{c|c|c}
    \multicolumn{3}{c}{Scalar Fields}\\
    \hline
    $\phi_{h}$ & $\phi_{H}$ & $\phi_{DM}$ \\
    \hline
    $0$ & $1$ & $n_{\mu \tau}$\\
\end{tabular}\\
\hline
\hline
\end{tabular}
\caption{Particle contents and their corresponding
charges under $\umt$.}
\label{tab2}
\end{table}
\end{center}
 
We now write the Lagrangian of present model, which is given by
\begin{eqnarray}
\mathcal{L}&=&\mathcal{L}_{SM} + \mathcal{L}_{N} + \mathcal{L}_{DM}
+ (D_{\mu}\phi_{H})^{\dagger} (D^{\mu}\phi_{H})
-V(\phi_{h},\phi_{H})
-\frac{1}{4} F_{\mu \tau}^{\alpha \beta} {F_{\mu \tau}}_{\alpha \beta}
\,, 
\label{lag}
\end{eqnarray}
where $\mathcal{L}_{SM}$ is the usual SM Lagrangian while
the Lagrangian for the right handed neutrinos containing
their kinetic energy terms, mass terms and Yukawa terms
with the SM lepton doublets, is denoted by $\mathcal{L}_{N}$
which can be written as
\begin{eqnarray}
\mathcal{L}_{N}&=&
\sum_{i=e,\,\mu,\,\tau}\frac{i}{2}\bar{N_i}\gamma^{\mu}D_{\mu} N_{i} 
-\dfrac{1}{2}\,M_{ee}\,\bar{N_e^{c}}N_{e}
-\dfrac{1}{2}\,M_{\mu \tau}\,(\bar{N_{\mu}^{c}}N_{\tau}
+\bar{N_{\tau}^{c}}N_{\mu})  \nn \\ &&
-\dfrac{1}{2}\,h_{e \mu}(\bar{N_{e}^{c}}N_{\mu} 
+\bar{N_{\mu}^{c}}N_{e})\phi_H^\dagger
- \dfrac{1}{2}\,h_{e \tau}(\bar{N_{e}^{c}}N_{\tau} 
+ \bar{N_{\tau}^{c}}N_{e})\phi_H
\nn \\ &&
-\sum_{i=e,\,\mu,\,\tau} y_{i} \bar{L_{i}}
\tilde {\phi_{h}} N_{i} +h.c.\,
\label{lagN}
\end{eqnarray}
with $\tilde {\phi_{h}}=i\,\sigma_2\phi^*_h$
and $M_{ee}$, $M_{\mu \tau}$ are constants having dimension
of mass while the Yukawa couplings $h_{e\mu}$, $h_{e \tau}$ and $y_i$ are
dimensionless constants. In Eq.\,(\ref{lag}), $\mathcal{L}_{DM}$
represents the dark sector Lagrangian including
the interactions of $\phi_{DM}$ with other
scalar fields. The expression of $\mathcal{L}_{DM}$ is given by
\begin{eqnarray}
\mathcal{L}_{DM} &=& (D^{\mu}\phi_{DM})^\dagger (D_{\mu}\phi_{DM})
- \mu_{DM}^{2} \phi_{DM}^{\dagger} \phi_{DM} 
-\lambda_{DM} (\phi_{DM}^{\dagger} \phi_{DM})^{2}
\nn \\ && 
-\lambda_{Dh}(\phi_{DM}^{\dagger} \phi_{DM})
(\phi_{h}^{\dagger} \phi_{h}) 
-\lambda_{DH}(\phi_{DM}^{\dagger} \phi_{DM})(\phi_{H}^{\dagger} \phi_{H})
 \,.
\label{ldm}
\end{eqnarray}
Moreover, the quantity $V(\phi_{h},\phi_{H})$ in Eq.\,(\ref{lag})
contains all the self interaction of $\phi_H$ and its
interaction with SM Higgs doublet. Therefore,
\begin{eqnarray}
V(\phi_h, \phi_H) = \mu_{H}^{2} \phi_{H}^{\dagger} \phi_{H} 
+ \lambda_{H} (\phi_{H}^{\dagger} \phi_{H})^{2}
+ \lambda_{hH}(\phi_{h}^{\dagger} \phi_{h}) (\phi_{H}^{\dagger} \phi_{H}) \,.
\label{int}
\end{eqnarray} 
The expressions of all the covariant derivatives appearing
in Eqs.\,(\ref{lag})-(\ref{ldm}) can be written in a generic form
which is given as
\begin{eqnarray}
D_{\nu} X = (\partial_{\nu} + i\,\gmt\,Q_{\mu \tau} (X)\,{\zmt}_{\nu})\,X\,,
\end{eqnarray}
where $X$ is any field which is singlet under SM gauge group
but has a $\lm-\lt$ charge $Q_{\mu \tau} (X)$ (see Table \ref{tab2})
and $\gmt$ is the gauge coupling of the $\umt$ group. Furthermore,
the last term in Eq.\,(\ref{lag}) represents the kinetic term
for the extra neutral gauge boson $\zmt$ in terms of its field strength
tensor $F_{\mu \tau}^{\alpha \beta} = \partial^\alpha \zmt^\beta
-\partial^\beta \zmt^\alpha.$

The $\lm-\lt$ symmetry breaks spontaneously when $\phi_H$ acquires
VEV and consequently the corresponding gauge field $\zmt$ becomes
massive, $M_{\zmt}=\gmt\,\vmt$. In the unitary gauge, the
expressions of $\phi_{h}$ and $\phi_{H}$ after spontaneous
breaking of the $\smgauge\times\umt$ gauge symmetry are
\begin{eqnarray}
\phi_{h}=
\begin{pmatrix}
0 \\
\dfrac{v+H}{\sqrt{2}}
\end{pmatrix}\,,
\,\,\,\,\,\,\,\,\,
\phi_{H}=
\begin{pmatrix}
\dfrac{\vmt + H_{\mu\tau}}{\sqrt{2}}
\end{pmatrix}\,,
\label{phih}
\end{eqnarray}
where $v$ and $\vmt$ are the VEVs of $\phi_h$ and $\phi_H$
respectively. Presence of the mutual interaction term in Eq.\,(\ref{int})
between $\phi_h$ and $\phi_H$ introduces mass mixing between the
scalar fields $H$ and $H_{\mu\tau}$. The scalar mass matrix
with off-diagonal elements proportional to $\lambda_{hH}$ is given by
\begin{eqnarray}
\mathcal{M}^2_{scalar} = \left(\begin{array}{cc}
2\lambda_h\,v^2 ~~&~~ \lambda_{hH}\,\vmt\,v \\
~~&~~\\
\lambda_{hH}\,\vmt\,v ~~&~~ 2 \lambda_H\,\vmt^2
\end{array}\right) \,\,.
\label{mass-matrix}
\end{eqnarray}
From the expression of $\mathcal{M}^2_{scalar}$
it is evident that if $\lambda_{hH}=0$ (i.e. the
interaction between $\phi_h$ and $\phi_H$ is absent),
there is no mixing between $H$ and $H_{\mu\tau}$
and hence they can represent two physical states. In our 
model however $\lambda_{hH} \neq 0$ and consequently the
states representing the physical scalars will be obtained
after the diagonalization of matrix $\mathcal{M}^2_{scalar}$.
The new physical states which are linear combinations of
$H$ and $H_{\mu\tau}$ can be written as
\begin{eqnarray}
h_{1}&=& H \cos \alpha + H_{\mu\tau} \sin \alpha \,, \nn \\
h_{2}&=& - H \sin \alpha + H_{\mu\tau} \cos \alpha\,.
\end{eqnarray}
The mixing angle $\alpha$ and the corresponding
eigenvalues (masses of $h_1$ and $h_2$) are given by
\begin{eqnarray}
\tan 2\alpha &=& \dfrac{\lambda_{hH}\,\vmt\,v}
{\lambda_h v^2 - \lambda_H \vmt^2}\,,
\label{scalarmix}\\
M^2_{h_1} &=& \lambda_h v^2 + \lambda_H \vmt^2 + 
\sqrt{(\lambda_h v^2 - \lambda_H \vmt^2)^2 + (\lambda_{hH}\,v\,\vmt)^2}
\ ,
\label{massh1} \\
M^2_{h_2} &=& \lambda_h v^2 + \lambda_H \vmt^2 - 
\sqrt{(\lambda_h v^2 - \lambda_H \vmt^2)^2 + (\lambda_{hH}\,v\,\vmt)^2}\,.
\label{massh2}
\end{eqnarray}
We have considered $h_1$ as the SM-like Higgs boson
\footnote{Eq.\,(\ref{massh1}, \ref{massh2}) are valid when $M_{h_1}>M_{h_2}$.
On the other hand, the expressions of $M_{h_1}$ and $M_{h_2}$
will be interchanged for $M_{h_2}>M_{h_1}$ resulting an change in sign
to the mixing angle $\alpha$.}
which has recently
been discovered by ATLAS \cite{Aad:2012tfa}
and CMS \cite{Chatrchyan:2012xdj} collaborations.
Therefore its mass $M_{h_1}$ and VEV $v$ are kept fixed at
125.5 GeV and 246 GeV respectively. The mass of dark
matter candidate $\dm$ takes the following form
\begin{eqnarray}
\mdm^2 = \mu^2_{DM} + \dfrac{\lambda_{Dh}\,v^2}{2} +
\dfrac{\lambda_{DH}\,\vmt^2}{2}\,.
\end{eqnarray}
In this model our ground state is defined as
$\langle\phi_{h}\rangle = \dfrac{v}{\sqrt{2}}$,
$\langle\phi_{H}\rangle = \dfrac{\vmt}{\sqrt{2}}$ and
$\langle\dm\rangle = 0$ this requires 
\begin{eqnarray}
\mu^2_h<0,\,\,\,\mu^2_H<0\,\,\text{and}\,\,\,\mu^2_{DM}>0.
\end{eqnarray}
The stability of the ground state (vacuum) requires the following
inequalities \cite{Biswas:2016ewm} among the quartic couplings of scalar fields  
\begin{eqnarray}
&&\lambda_h \geq 0, \lambda_H \geq 0, \lambda_{DM} \geq 0,\nonumber \\
&&\lambda_{hH} \geq - 2\sqrt{\lambda_h\,\lambda_H},\nonumber \\
&&\lambda_{Dh} \geq - 2\sqrt{\lambda_h\,\lambda_{DM}},\nonumber \\
&&\lambda_{DH} \geq - 2\sqrt{\lambda_H\,\lambda_{DM}},\nonumber \\
&&\sqrt{\lambda_{hH}+2\sqrt{\lambda_h\,\lambda_H}}\sqrt{\lambda_{Dh}+
2\sqrt{\lambda_h\,\lambda_{DM}}}
\sqrt{\lambda_{DH}+2\sqrt{\lambda_H\,\lambda_{DM}}} \nonumber \\ 
&&+ 2\,\sqrt{\lambda_h \lambda_H \lambda_{DM}} + \lambda_{hH} \sqrt{\lambda_{DM}}
+ \lambda_{Dh} \sqrt{\lambda_H} + \lambda_{DH} \sqrt{\lambda_h} \geq 0 \,\,\,\,.
\end{eqnarray} 
Besides the above inequalities, the upper bound on quartic, gauge
and Yukawa couplings can be obtained from the condition of
perturbativity. For a scalar quartic coupling $\lambda$
($\lambda=\lambda_{h}$,\, $\lambda_H$,\, $\lambda_{DM}$,\,
$\lambda_{hH}$,\, $\lambda_{Dh}$,\, $\lambda_{DH}$)
this condition will be ensured when \cite{Chakrabarty:2015yia}
\begin{eqnarray}
\lambda<4\pi\,, 
\label{par_quartic}
\end{eqnarray}
while for gauge coupling $\gmt$ and Yukawa coupling $y$ ($y=y_e$,
$y_\mu$, $y_\tau$, $h_{e\mu}$ and $h_{e\tau}$) it is
\cite{Chakrabarty:2015yia} 
\begin{eqnarray}
\gmt,\,y<\sqrt{4\pi}\,. 
\label{par_gauge}
\end{eqnarray}
The above quadratic and quartic couplings of scalars fields $\phi_h$
and $\phi_H$ namely $\mu^2_h$, $\mu^2_H$, $\lambda_{h}$, $\lambda_{H}$
and $\lambda_{hH}$ can be expressed in terms of physical
scalar masses ($M_{h_1}$, $M_{h_2}$), mixing angle $\alpha$
and VEVs ($v$, $\vmt$), which have been given in \cite{Biswas:2016ewm}.
\section{Muon $(g-2)$}
\label{muong-2}
\begin{figure}[h!]
\centering
\includegraphics[angle=0,height=6cm,width=8cm]{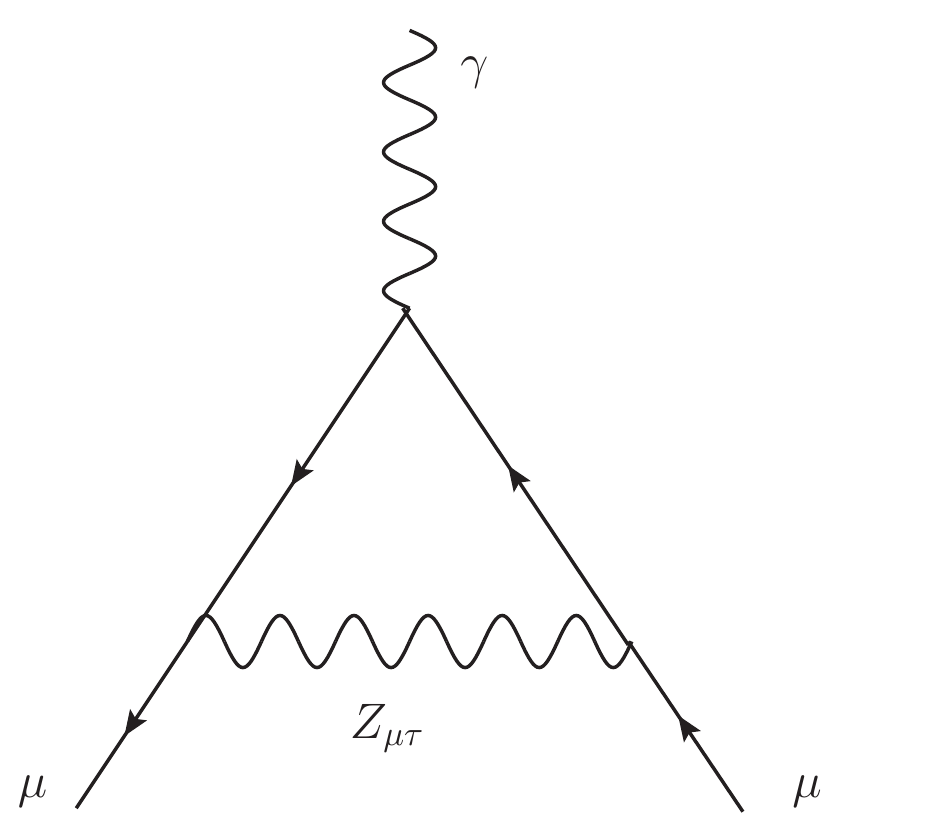}
\caption{One loop Feynman diagram contributing to muon $(g-2)$, 
mediated by the extra gauge boson $\zmt$. }
\label{muong2}
\end{figure}

It is well known that from the Dirac equation, the magnetic
moment of muon $\vec{M}$ can be written in terms of its spin ($\vec{S}$),
which is
\begin{eqnarray}
\vec{M} = g_{\mu} \dfrac{e}{2\,m_\mu} \vec{S},
\end{eqnarray}
where $m_{\mu}$ is the mass of muon and $g_{\mu}=2$ is the
gyromagnetic ratio. However, if we calculate $g_{\mu}$
using QFT then contributions arising from loop corrections
slightly shift the value of $g_{\mu}$ from 2. Hence one can
define a quantity $a_{\mu}$ which describes the deviation of
$g_{\mu}$ from its tree level value,
\begin{eqnarray}
a_{\mu} = \dfrac{g_{\mu}-2}{2}\,.
\end{eqnarray}
In general, the contribution to the theoretical value of
$a_{\mu}$ ($a_{\mu}^{\rm th}$) comes from the following
sources \cite{Agashe:2014kda}
\begin{eqnarray}
a_{\mu}^{\rm th} = a_{\mu}^{\rm QED} + a_{\mu}^{\rm EW}
+ a_{\mu}^{\rm Had}\,, 
\end{eqnarray} 
where the contributions arising from Quantum Electrodynamics (QED), 
Electroweak theory and hadronic process are denoted by
${\rm a_{\mu}^{QED}}$, ${\rm a_{\mu}^{EW}}$ and
${\rm a_{\mu}^{Had}}$, respectively. The SM prediction
of $a_{\mu}$ including the above terms is \cite{Jegerlehner:2009ry}
\begin{eqnarray}
a_{\mu}^{\rm th} = 1.1659179090(65)\times 10^{-3}\,.
\end{eqnarray}
On the other hand, 
$a_\mu$ has been precisely measured experimentally, 
initially by the CERN experiments
and later on by the E821 experiment, and the current average experimental
value is \cite{Bennett:2004pv} 
\begin{eqnarray}
a_{\mu}^{\rm exp} = 1.16592080(63) \times 10^{-3}\,.
\end{eqnarray}
From the above one can see that although the theoretically
predicted and the experimentally measured values of $a_{\mu}$
are quite close to each other, there still exists
some discrepancy between these two quantities at the $3.2\sigma$
significance which is  \cite{Jegerlehner:2009ry},
\begin{eqnarray}
\Delta a_{\mu} = a_{\mu}^{\rm exp} - a_{\mu}^{\rm th}
= (29.0 \pm 9.0) \times 10^{-10}\,.
\label{delta}
\end{eqnarray}
Therefore, in order to reduce the difference between $a_{\mu}^{\rm exp}$
and $a_{\mu}^{\rm th}$ we need to explore BSM scenarios where we can get
extra contributions from some extra diagrams. In our
$\umt$ model we have an additional one loop diagram compared to
the SM, which is mediated by the extra neutral gauge boson $\zmt$
and gives nonzero contribution to $a_{\mu}^{\rm th}$
as shown in Fig.\,\ref{muong2}.
The additional contribution to $a_{\mu}^{\rm th}$  from this diagram
 is given by \cite{Gninenko:2001hx, Baek:2001kca},
\begin{eqnarray}
\Delta a_{\mu}(Z_{\mu \tau}) = \dfrac{g_{\mu \tau}^{2}}
{8 \pi^{2}} \int_{0}^{1} dx \dfrac{2 x(1-x)^{2}}{(1-x)^{2} + rx}\,,
\label{intg2}
\end{eqnarray}
where, $r = (M_{Z_{\mu \tau}}/m_{\mu})^{2}$ is the square of the ratio
between masses of gauge boson ($\zmt$) and muon. As mentioned
in the Introduction, although
a $\mathcal{O}$(100 MeV) $\zmt$ is allowed, its coupling strength ($\gmt$)
is strongly constrained to be less than $\sim 10^{-3}$ from the measurement of
neutrino trident cross section by experiments like
CHARM-II \cite{Geiregat:1990gz} and CCFR \cite{Mishra:1991bv}. 
In our analysis, we find that for $\mzmt=100$ MeV and
$\gmt=9 \times 10^{-4}$ the value of $\Delta a_{\mu} = 22.6 \times 10^{-10}$,
which lies around the ballpark value given in
Eq.\,(\ref{delta}). In what follows, we will use 
$\mzmt=100$ MeV and $\gmt=9.0\times10^{-3}$ as
our benchmark point for the analyses of neutrino masses 
and dark matter phenomenology. 
\section{Neutrino Masses and Mixing}
\label{numass}
Majorana neutrino masses are generated via the Type-I seesaw 
mechanism by
the addition
of three right handed neutrinos to the model. Using Eq.\,(\ref{lagN})
we can write the Majorana mass matrix for the three right handed
neutrinos as
\begin{eqnarray}
\mathcal{M}_{R} = \left(\begin{array}{ccc}
M_{ee} ~~&~~ \dfrac{ \vmt}{\sqrt{2}} h_{e \mu}
~~&~~\dfrac{\vmt}{\sqrt{2}} h_{e \tau} \\
~~&~~\\
\dfrac{\vmt}{\sqrt{2}} h_{e \mu} ~~&~~ 0
~~&~~ M_{\mu \tau} \,e^{i\xi}\\
~~&~~\\
\dfrac{\vmt}{\sqrt{2}} h_{e \tau} ~~&
~~ M_{\mu \tau}\,e^{i\xi} ~~&~~ 0 \\
\end{array}\right) \,,
\label{mncomplex}
\end{eqnarray} 
where all parameters in $M_R$ 
in general can be complex. However, by proper phase rotation
one can choose all the elements expect the ${\mu\tau}$ component of 
$M_R$ to be real \cite{Baek:2015mna}. Thus, $M_R$ depends on the real parameters 
$M_{ee}$, $M_{\mu\tau}$, $h_{e\mu}$ and $h_{e\tau}$ and the phase $\xi$. 
On other hand, from the Yukawa term in Eq.\,(\ref{lagN})
one can easily see that the Dirac mass matrix $M_D$ between
left handed and right handed neutrinos is
diagonal and for simplicity we have chosen all the Yukawa
couplings ($y_e$, $y_{\mu}$ and $y_{\tau}$) are real. The
expression of $M_D$ is
\begin{eqnarray}
M_{D} = \left(\begin{array}{ccc}
f_e ~~&~~ 0 ~~&~~ 0 \\
~~&~~\\
0 ~~&~~ f_{\mu} ~~&~~ 0 \\
~~&~~\\
0 ~~&~~ 0 ~~&~~ f_{\tau} \\
\end{array}\right) \,,
\label{md}
\end{eqnarray}
where $f_i = \dfrac{y_{i}}{\sqrt{2}}v$ with $i=e$, $\mu$ and $\tau$.
Now, with respect to the basis $\left(\overline{{\nu_\alpha}_L}
~~\overline{({N_{\alpha}}_R)^c}\right)^T$ and $\left(({\nu_\alpha}_L)^c
~~{N_{\alpha}}_R\right)^T$ we can write the mass matrix
of both left as well as right handed neutrinos which
is given as
\begin{eqnarray}
M = \left(\begin{array}{cc}
0 & M_D \\
M^T_D & M_R
\end{array}\right) \,\,,
\label{mtot}
\end{eqnarray}
where $M$ is a $6\times6$ matrix and both $M_D$
and $M_R$ are $3\times 3$ matrices given by Eqs. (\ref{mncomplex}) and (\ref{md}). 
After diagonalisztion
of the matrix $M$ one obtains two fermionic states
for each generation which are Majorana in nature.
Therefore we have altogether six Majorana
neutrinos, out of which three are light and
rest are heavy. Using block
diagonalisation technique, we can find
the mass matrices for light 
as well as heavy neutrinos which are given as 
\begin{eqnarray}
m_{\nu}&\simeq&-M_D\,M^{-1}_R M^T_D\,, 
\label{activemass}\\
m_N &\simeq& M_R\,.
\label{sterilemass}
\end{eqnarray}  
Here both $m_{\nu}$ and $m_N$ are complex
symmetric matrices. Also
Eqs.\,(\ref{activemass}-\ref{sterilemass}) are derived
using an assumption that $M_D\ll M_R$ i.e. the eigenvalues
of $M_D$ is much less than those of $M_R$ and therefore
terms with higher powers of $M_{D}/M_R$ are neglected. Using
the expressions of $M_R$ and $M_D$ given in Eqs.\,(\ref{mncomplex}-\ref{md})
the light neutrino mass matrix in this model
takes the following form
\begin{eqnarray}
m_{\nu} = \dfrac{1}{2\,p} \left(\begin{array}{ccc}
2\,f_{e}^{2}M_{\mu \tau}^{2}e^{i\xi} &
-\sqrt{2}\,f_{e} f_{\mu}\,h_{e \tau} \vmt &
-\sqrt{2}\,f_{e} f_{\tau}\,h_{e \mu} \vmt\\
-\sqrt{2}\,f_{e} f_{\mu}\,h_{e \tau} \vmt &
\dfrac{f_{\mu}^{2}\,h_{e\tau}^2\,\vmt^2\,e^{-i\xi}}{M_{\mu\tau}} &
\dfrac{f_{\mu}\,f_{\tau}}{M_{\mu\tau}}(M_{ee}\,M_{\mu\tau}-p\,e^{-i\xi})\\
-\sqrt{2}\,f_{e} f_{\tau}\,h_{e \mu} \vmt &
\dfrac{f_{\mu}\,f_{\tau}}{M_{\mu\tau}}(M_{ee}\,M_{\mu\tau}-p\,e^{-i\xi})  &
\dfrac{f_{\tau}^{2}\,h_{e\mu}^2\,\vmt^2\,e^{-i\xi}}{M_{\mu\tau}} \\
\end{array}\right) \,\,,
\label{mass-matrix}
\end{eqnarray}
where $p = h_{e\mu}\,h_{e\tau}\,\vmt^2-M_{ee}\,M_{\mu\tau}\,e^{i\xi}$.
The masses and mixing angles of the light neutrinos are
found by diagonalising this matrix \cite{Adhikary:2013bma}
and are compared against the corresponding experimentally
allowed ranges obtained from global analysis of the data
(cf. Eq. (\ref{eq:mix})). 

There are eight independent parameters in the light neutrino mass matrix
$m_{\nu}$, namely, $f_{e}$, $f_{\mu}$, $f_{\tau}$, $M_{\mu\tau}$, $M_{ee}$,
$V_{e\tau}=\frac{\vmt}{\sqrt{2}}\,h_{e\tau}$,
$V_{e\mu}=\frac{\vmt}{\sqrt{2}}\,h_{e\mu}$ and $\xi$.
All of these parameters have mass dimension GeV
except the dimensionless phase factor $\xi$ which is
in radian. In order to find the model parameter space allowed
by the neutrino oscillation experiments,
we have varied the above mentioned parameters in the
following range
\begin{eqnarray}
\begin{array}{cccccc}
0 &\le & \xi\,[\text{rad}]& \le& 2\pi\,\,,\\
1 & \le & M_{ee},\,M_{\mu\tau}\,[\text{GeV}]& \le & 10^4\,\, ,\\
1 &\le & V_{e\mu},\,\,\,V_{e\tau}\,\,\,[\text{GeV}]& \le & 280\,\, ,\\
0.1&\le & \dfrac{(f_{e},\,f_{\mu},\,f_{\tau})}{10^{-4}}\,[\text{GeV}]
& \le& 10\,\, .
\label{para-ranges}
\end{array}
\end{eqnarray}
The allowed parameter space satisfies the following constraints from the neutrino 
sector
\begin{itemize}
\item cosmological upper bound on the sum of all three light neutrinos,
$\sum_i m_{i} < 0.23$ eV at $2\sigma$ C.L. \cite{Ade:2015xua},
\item mass squared differences $6.93<\dfrac{\Delta m^2_{21}}
{10^{-5}}\,{\text{eV}^2} < 7.97$ and $2.37<\dfrac{\Delta m^2_{31}}
{10^{-3}}\,{\text{eV}^2} < 2.63$ in $3\sigma$ range \cite{Capozzi:2016rtj},
\item all three mixing angles $30^{\circ}<\,\theta_{12}\,<36.51^{\circ}$,
$37.99^{\circ}<\,\theta_{23}\,<51.71^{\circ}$ and
$7.82^{\circ}<\,\theta_{13}\,<9.02^{\circ}$ also in $3\sigma$ range
\cite{Capozzi:2016rtj}.
\end{itemize} 
All the Yukawa couplings appearing in the light as well as heavy
Majorana neutrino mass matrices ($m_{\nu}$ and $M_R$) are enforced to always lie within
the perturbative range mentioned in Eq.\,(\ref{par_gauge}). 
Furthermore, we scan the allowed areas in the model parameter space 
for only for the normal mass ordering which corresponds to $\Delta m_{31}^2 > 0$. 
\begin{figure}[h!]
\centering
\includegraphics[angle=0,height=7.5cm,width=8.0cm]{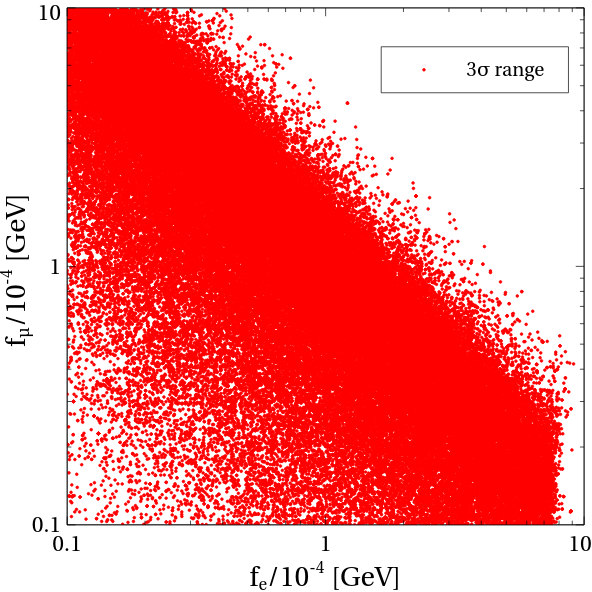}
\includegraphics[angle=0,height=7.5cm,width=8.0cm]{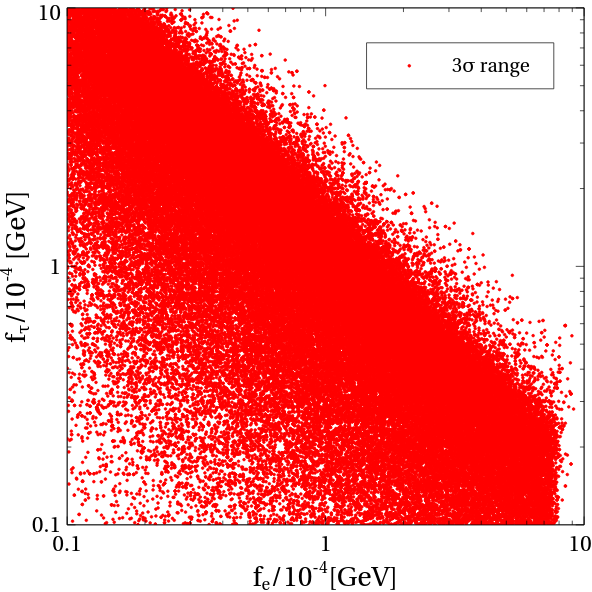}
\caption{Left (Right) panel: Allowed region in $f_e-f_{\mu}$
($f_e-f_{\tau}$) plane which satisfies all the experimental
constraints considered in this work.}      
\label{a1}
\end{figure}

In the left and right panels of Fig.\,\ref{a1}, we have shown the allowed
regions in $f_{e}-f_{\mu}$ and $f_{e}-f_{\tau}$ planes respectively, 
where we have varied $f_{e}$, $f_{\mu}$, $f_{\tau}$ in the range $10^{-5}$ GeV
to $10^{-3}$ GeV while the other parameters have been scanned over the
entire considered range as given in Eq.\,(\ref{para-ranges}). From both
the panels it is clear that there is (anti)correlation between the parameters
$f_{e}-f_{\mu}$ and $f_{e}-f_{\tau}$. We find that for the lower
values of $f_{e}$ higher values of $f_{\mu}$, $f_{\tau}$ are needed
to satisfy the experimental constraints in the $3\sigma$ range and vice versa.
Moreover, although there are smaller number of allowed points
when both $f_{e}$ and $f_i$ ($i=\mu$, $\tau$) are small but the present
experimental bounds on the observables of the neutrino sector
forbid the entire region in the 
$f_{e}-f_{\mu}$ and  $f_{e}-f_{\tau}$ planes for both $f_e$ and
$f_i>2\times 10^{-4}$ GeV ($i=\mu$, $\tau$). Also, unlike
the parameters $f_{\mu}$ and $f_{\tau}$,
we do not get any allowed values of $f_e$ beyond $8\times 10^{-4}$ GeV.
\begin{figure}[h!]
\centering
\includegraphics[angle=0,height=7.5cm,width=8.0cm]{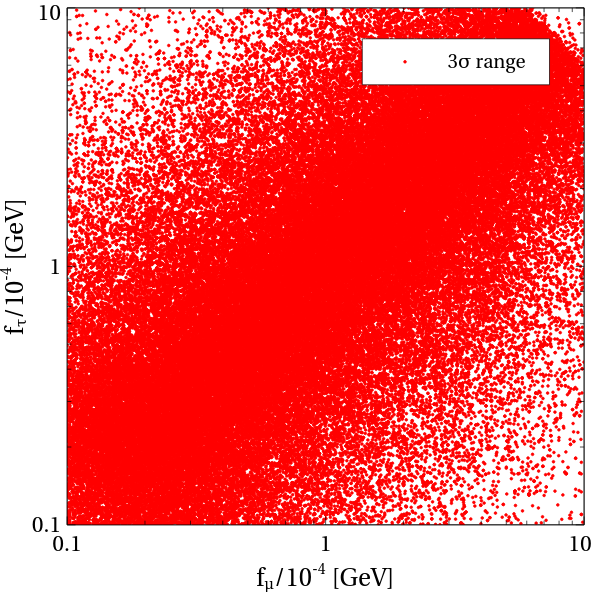}
\includegraphics[angle=0,height=7.50cm,width=8.0cm]{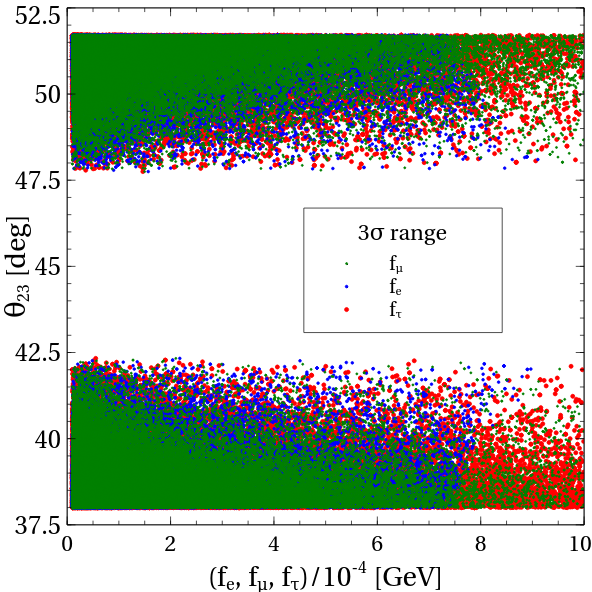}
\caption{Left panel: Allowed region in $f_{\mu}-f_{\tau}$ plane.
Right panel: Variation of $\theta_{23}$ with $f_{e}$ (blue dots),
$f_{\mu}$ (green dots) and $f_{\tau}$ (red dots).}      
\label{b1}
\end{figure}

The allowed parameter space in $f_{\mu}-f_{\tau}$ plane has been
shown in the left panel of Fig.\,\ref{b1}. From the figure
it is seen that there is a correlation between the
parameters $f_{\mu}$ and $f_{\tau}$. That means unlike the
previous plots here most of allowed points in $f_{\mu}-f_{\tau}$
plane are such that for the lower (higher) values of the parameter
$f_{\mu}$ we also need lower (higher) values of $f_{\tau}$ to
reproduce the experimental results. On the other hand,
in the right panel of Fig.\,\ref{b1}, we show the variation
of $\theta_{23}$ with $f_{e}$ (blue dots),
$f_{\mu}$ (green dots) and $f_{\tau}$ (red dots). 
We see from the plot that the region around maximal $\theta_{23}$ 
mixing angle is ruled out in this model. The reason is that while in the 
$\lm-\lt$ symmetric limit, the neutrino mass matrix had a $\mu-\tau$ symmetry and 
hence $\theta_{23}=\pi/4$ and $\theta_{13}=0$, once the $\lm-\lt$ symmetry is 
spontaneously broken, $\theta_{23}$ shifts away from maximal and $\theta_{13}$ 
becomes non-zero, making the model consistent with the neutrino oscillations data.  
The plot also shows that the allowed values of
mixing angle $\theta_{23}$ lie in two separate ranges between
$38^{\circ}\la\,\theta_{23}\,\la 42^{\circ}$
(lower octant, $\theta_{23}<$ $45^\circ$) and $48^{\circ}\la\,
\theta_{23}\,\la 51.5^{\circ}$
(higher octant, $\theta_{23}>45^\circ$)
for the variation of entire considered range of parameters
$f_i$ ($i=e,\,\mu,\,\tau$) from $10^{-5}$ GeV to $10^{-3}$
GeV. Therefore, we can conclude that our model is
insensitive to the octant of $\theta_{23}$. 
\begin{figure}[h!]
\centering
\includegraphics[angle=0,height=7.3cm,width=8.5cm]{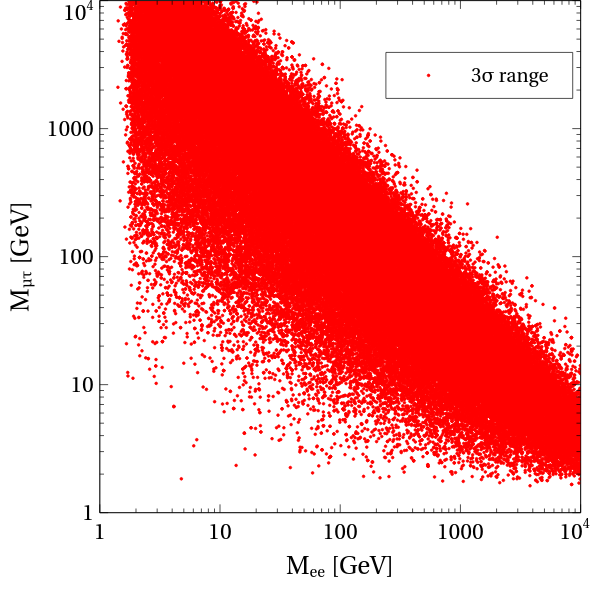}
\includegraphics[angle=0,height=7.5cm,width=8.0cm]{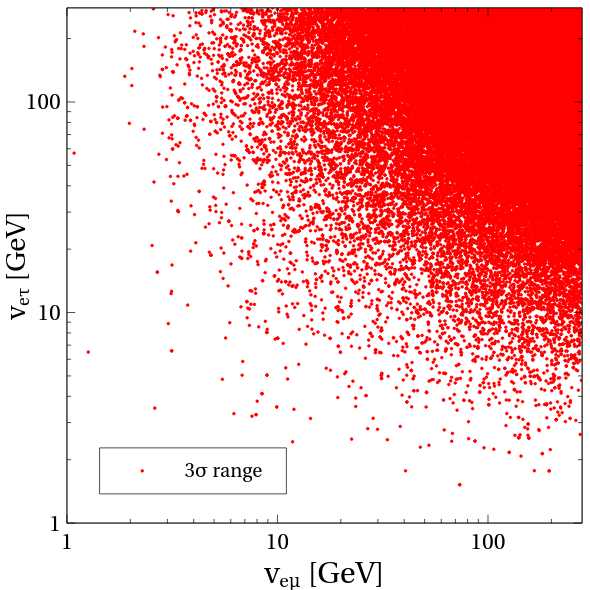}
\caption{Left (Right) panel: Allowed region in $M_{ee}-M_{\mu\tau}$
($V_{e\mu}-V_{e\tau}$) plane which satisfies all the experimental
constraints considered in this work.}      
\label{c1}
\end{figure}

The allowed regions for the other remaining parameters
$M_{ee}-M_{\mu\tau}$ and $V_{e\mu}-V_{e\tau}$ have been
shown in Fig.\,\ref{c1}. The left panel of
Fig.\,\ref{c1} shows the (anti)correlation
between the allowed values of the parameters $M_{ee}$ and $M_{\mu \tau}$.
The neutrino oscillation data rules out the parameter region 
$M_{ee}\ga 500$ GeV, $M_{\mu \tau} \ga 500$ GeV and 
$M_{ee}\la 5$ GeV, $M_{\mu \tau} \la 5$ GeV. 
In the right panel Fig.\,\ref{c1}, we have shown the allowed
region in the $V_{e\mu}-V_{e\tau}$ plane. In order to keep the
Yukawa couplings $h_{e\mu}$ and $h_{e\tau}$ within the perturbative
regime (see Eq.\,(\ref{par_gauge})) we have restricted variation
of both $V_{e\mu}$ and $V_{e\tau}$ upto 280 GeV.
From this plot it is clearly seen that the higher
values of $V_{e\mu}$ and $V_{e\tau}$
($V_{e\mu},\,V_{e\tau}\ga10$ GeV) are mostly preferred by the
neutrino experiments over the smaller ones.
\begin{figure}[h!]
\centering
\includegraphics[angle=0,height=7.5cm,width=8.0cm]{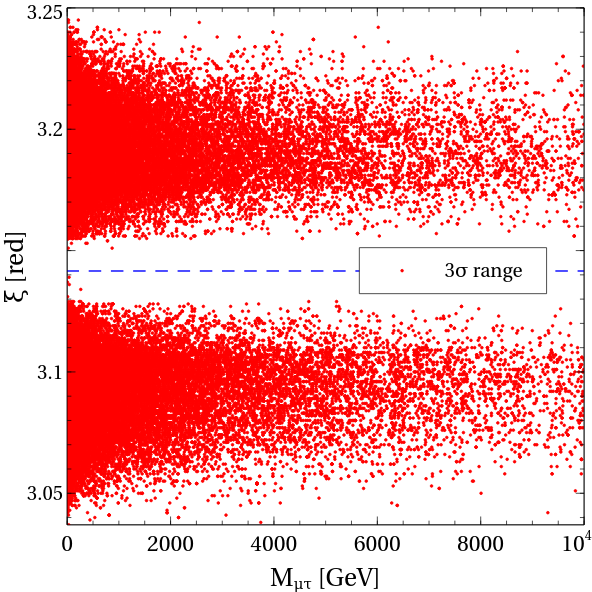}
\includegraphics[angle=0,height=7.5cm,width=9.1cm]{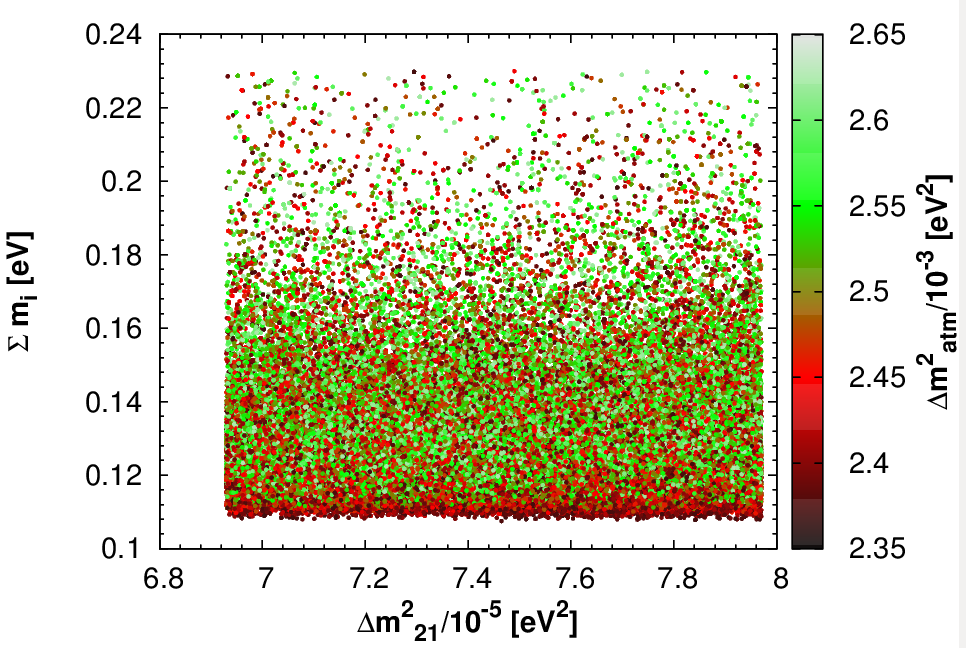}
\caption{Left pane: Allowed values of the parameters
$M_{\mu\tau}$ and $\xi$. Blue dashed line represents $\xi=\pi$.
Right panel: Variation of $\sum_i m_{\nu_i}$
with the mass square differences $\Delta m^2_{21}$
and $\Delta m^2_{32}$.}      
\label{d1}
\end{figure}

In the left panel of Fig.\,\ref{d1}, we have shown the variation
of the phase $\xi$ with respect to the parameter $M_{\mu\tau}$. Only
a very narrow range of value of $\xi$, placed symmetrically with
respect to the line $\xi=\pi$, are allowed, which reproduce the
neutrino observables in the 3$\sigma$ range. It is also seen from
this figure that there are no points along $\xi=\pi$ line (blue dashed line),
which indicates that for the present model,
at least one element in the right handed neutrino
mass matrix (here we have considered $2\times3$ element of $M_R$) 
has to be a complex number to satisfy the experimental results.
The variation of sum of all three neutrino
masses with $\Delta m^{2}_{21}$
is presented in the right panel of Fig.\,\ref{d1}.
The variation of $\Delta m^{2}_{{\rm atm}}$ is also shown in the same figure.
From this plot, it is evident that in this model 
lower values of $\sum m_{i}$ ($\sum m_{i} \leq 0.18$ eV) are
more favourable.

\begin{figure}[h!]
\centering
\includegraphics[angle=0,height=7.5cm,width=8.0cm]{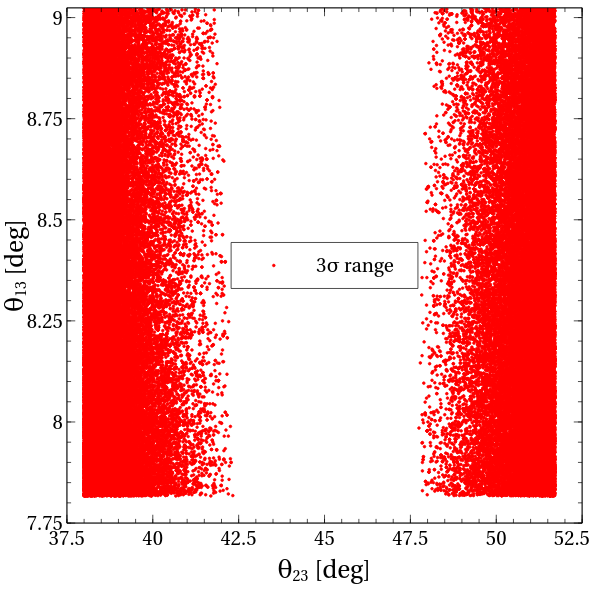}
\includegraphics[angle=0,height=7.5cm,width=8.0cm]{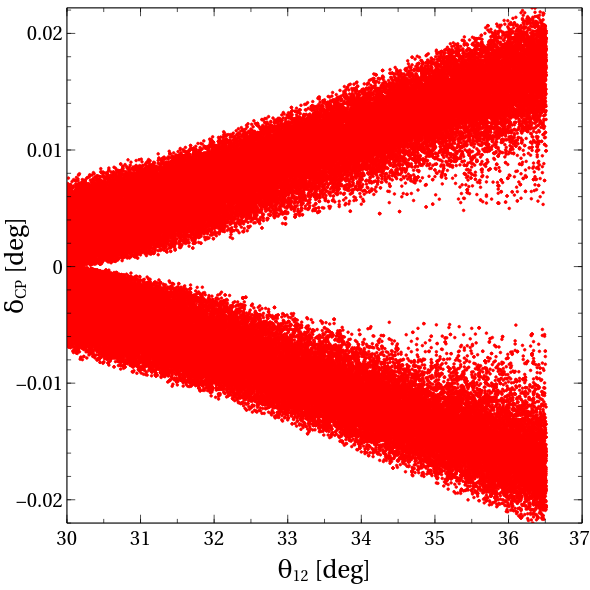}
\caption{Left panel: Variation of $\theta_{13}$ with $\theta_{23}$.
Right panel: Variation of Dirac CP phase $\delta_{CP}$ with mixing angle
$\theta_{12}$.}      
\label{e1}
\end{figure}

In the left and right panels of Fig.\,\ref{e1},
we have shown the predicted ranges of the mixing angles
and the Dirac CP phase. The left panel shows 
that for both lower and higher octant, the whole range of $\theta_{13}$
is allowed here. In the right panel of Fig.\,\ref{e1},
we have plotted the predicted Dirac CP phase
with respect to the mixing
angle $\theta_{12}$. We find that in our 
model the predicted values of Dirac CP phase are very small and   
symmetric around $0^\circ$. One can also note that 
the absolute predicted value of $|\delta_{CP}|$ increases with the 
mixing angle $\theta_{12}$. 
\section{Dark Matter}
\label{dm}
Being stable as well as electrically neutral,
$\dm$ can serve as a dark matter candidate. In this section,
we will compute the relic abundance of $\dm$ at the present epoch and
its spin independent scattering cross section relevant for direct detection experiments.
The viability of $\dm$ as a dark matter candidate will be tested
by comparing its relic abundance and spin independent scattering
cross section with the results obtained from Planck and
LUX experiments. Finally, at the end of this section we will
compute the $\gamma$-ray flux due to the annihilation of $\dm$
and compare this flux with Fermi-LAT observed $\gamma$-ray
excess from the regions close to the Galactic Centre (GC). 
\subsection{Relic Density}
\label{sec-RD}
\begin{figure}[h!]
\centering
\includegraphics[angle=0,height=4.5cm,width=13.5cm]{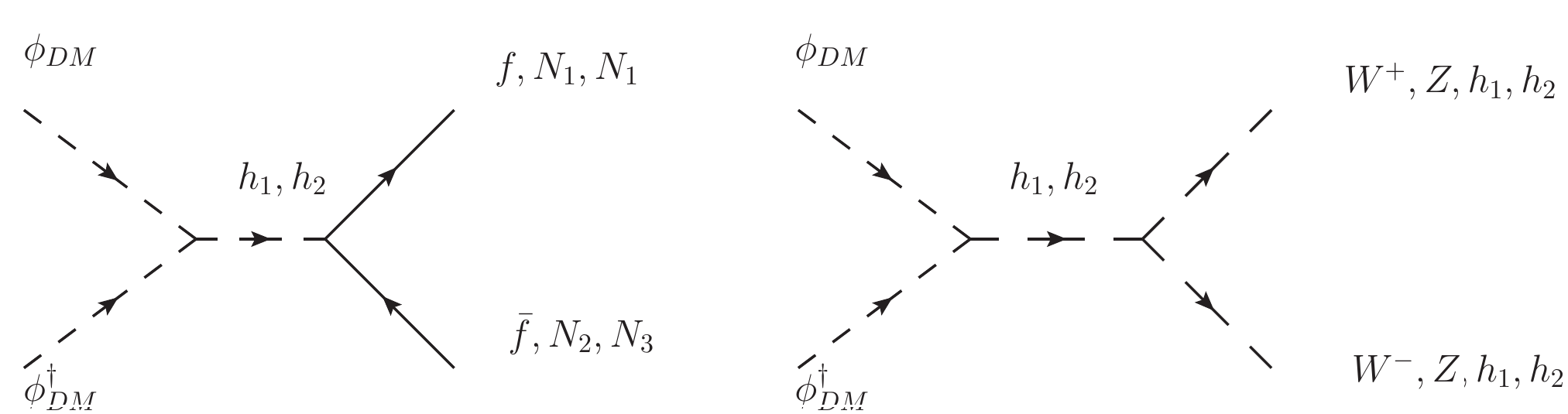}
\caption{Feynman diagrams dominantly contributing to the
annihilation cross section and hence towards the relic density of
$\dm$ and $\dmd$.}      
\label{feynann}
\end{figure}

In the present model, since $\dm$ is a complex scalar field
with a nonzero $\lm-\lt$ charge $\nmt$, therefore we have a
non-self-conjugate DM scenario where DM particle and its antiparticle
are different with respect to $\nmt$. In this work we assume
that there is no asymmetry between the number densities of
$\dm$ and $\dmd$ in the early Universe. The evolution of
total DM number density $n$ ($n=n_{\dm}$ + $n_{\dmd}$)
is governed by the well known Boltzmann
equation which is given by \cite{Gondolo:1990dk}
\begin{eqnarray}
\frac{dn}{dt} + 3 n{\rm H} = 
-\dfrac{1}{2}\langle{\sigma {\rm{v}}}\rangle 
\left(n^2 -n_{eq}^2\right) \,,
\label{be}
\end{eqnarray}
where $n_{eq}$ is the sum of equilibrium number densities
of both $\dm$, $\dmd$ and $H$ is the Hubble parameter. Moreover,
$\langle {\sigma\,{\rm v}} \rangle$ is the thermally
averaged annihilation cross section between
$\phi_{DM}$ and $\phi_{DM}^{\dagger}$ for the processes shown in 
Fig.\,\ref{feynann} \footnote{We have not shown $\zmt$ mediated diagrams as
the coupling strength of $\zmt$ with $\dm$ and $\dmd$ is proportional
to $\gmt$ which is needed to be very small ($\sim 10^{-3}$) for
the explanation of muon (${g-2}$) anomaly (see Section \ref{muong-2}).}.
In this work, we have considered DM mass in the range 30 GeV
to 500 GeV. Therefore depending on the value of $\mdm$,
$\dm$ and $\dmd$ can annihilate into the following final states:
$\dm \dmd \rightarrow f \bar{f}$, $W^+W^-$, $ZZ$, $\zmt\zmt$,
$h_1h_1$, $h_2h_2$, $h_1h_2$, $N_1\bar{N}_2$ and $N_1\bar{N}_3$ where $f$
is any SM fermion. The expressions of $\langle {\sigma\,{\rm v}} \rangle$ involving
actual annihilation cross section $\sigma$ 
and modified Bessel functions is given in \cite{Gondolo:1990dk}.
The factor $1/2$ appearing in the right hand side of the Boltzmann
equation is due to the non-self-conjugate nature of DM \cite{Gondolo:1990dk}.
In terms of two dimensionless quantities $Y$ and $x$ the above equation
can be written in the following form
\begin{eqnarray}
\frac{dY}{dx} =
-\left(\frac{45G}{\pi}\right)^{-\frac{1}{2}}
\frac{M_{DM}\,\sqrt{g_\star}}{x^2}\,
\dfrac{1}{2}\langle{\sigma {\rm{v}}}\rangle
\left(Y^2-(Y^{eq})^2\right)\,,
\label{be1}
\end{eqnarray}
where $Y=\frac{n}{\rm s}$ is the total comoving number density
of $\dm$ and $\dmd$ and $x=\frac{\mdm}{T}$ where $T$ is the temperature
of the Universe. Also, Newton's gravitational constant is denoted by $G$
while $g_{\star}$ is a function of effective degrees of freedom corresponding
to both energy and entropy densities of the Universe \cite{Gondolo:1990dk}. 
Therefore, the relic density of $\dm$ and $\dmd$ at the present epoch is
given by \cite{Edsjo:1997bg, Biswas:2011td}
\begin{eqnarray}
\Omega_{DM} h^2 = 2.755\times 10^8 \left(\frac{\mdm}
{\rm GeV}\right) Y(T_0)\,.
\end{eqnarray}
$Y(T_0)$ is the total comoving number density of $\dm$ and $\dmd$
for the present temperature of the Universe ($T_0\sim 10^{-13}$ GeV), 
which can be obtained by solving Eq.\,(\ref{be1}). 
\subsection{Direct detection}
\label{sec-DD}
Dark matter direct detection experiments use the principle of
elastic scattering between dark matter particles and
detector nuclei. If DM particles scatter off the
detector nuclei elastically then the information
about the nature of DM particles and their interaction type
with SM particles (quarks) can be obtained by
measuring the recoil energy of the nuclei. 
Since the DM particles are nonrelativistic (cold dark matter),
therefore the energy deposited to the nuclei
are extremely small ($\sim$ keV range).
Hence in order to measure it accurately, 
low background as well as low threshold
detector is required. In the present model,
the elastic scattering of both $\dm$ and $\dmd$
can occur only through the exchange of scalar bosons
$h_1$, $h_2$. Unlike the other U(1) extensions
of the SM where the extra neutral gauge bosons
can interact with the quarks (such as U(1)$_{B-L}$ model \cite{Biswas:2016ewm}),
here $\zmt$ does not couple with the quark sector and 
consequently, the spin independent
scattering cross sections of the DM particle and
its antiparticle are equal.
\begin{figure}[h!]
\centering
\includegraphics[angle=0,height=4.5cm,width=10cm]{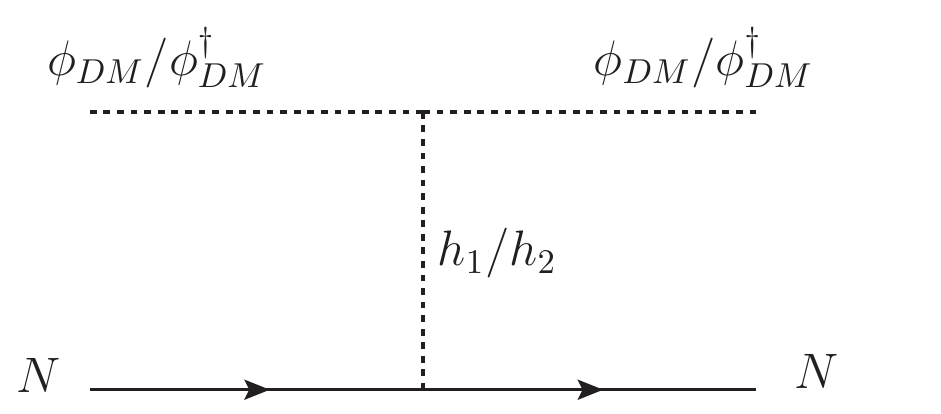}
\caption{Feynman diagram for the elastic scattering
of $\dm$ and $\dmd$ with detector nucleon ($N$).}      
\label{feyn}
\end{figure}
The expression of spin independent scattering cross section
of DM with nucleon ($N$) is given by
\begin{eqnarray}
\sigma_{\rm SI}
= \dfrac{\mu^2}{4\pi}\Bigg[\frac{M_N\,f_N\,\cos \alpha}{M_{DM}\,v}
\Bigg(\frac{\tan \alpha\,g_{\phi_{DM}\phi^{\dagger}_{DM} h_2}}{M^2_{h_2}}
-\frac{g_{\phi_{DM}\phi^{\dagger}_{DM} h_1}}{M^2_{h_1}}
\Bigg)\Bigg]^2,\nonumber\\
\label{sigmasi}
\end{eqnarray}
where $\mu$ is the reduced mass between DM and $N$ while
$f_N\sim0.3$ \cite{Cline:2013gha} is the nuclear form factor. 
$g_{\dm \dmd h_i}$ is the vertex factor involving fields $\dm$, $\dmd$
and $h_i$ ($i=1$, 2) and its expression is given in Table\,\ref{tab3}.
\begin{table}[h!]
\begin{center}
\vskip 0.5cm
\begin{tabular} {||c||c||}
\hline
\hline
Vertex & Vertex Factor\\
$a\,b\,c$ & $g_{abc}$\\
\hline
$q\,\bar{q}\,h_1$ & $-\dfrac{M_{q}}{v} \cos \alpha$\\
\hline
$q\,\bar{q}\,h_2$ & $ \dfrac{M_{q}}{v} \sin \alpha$\\
\hline
$W^{+}\,W^{-}\,h_1$ & $\dfrac{2\,M_{W}^{2}\cos \alpha}{v}$\\
\hline
$W^{+}\,W^{-}\,h_2$ & $-\dfrac{2\,M_{W}^{2}\sin \alpha}{v}$\\
\hline
$Z\,Z\,h_1$ & $\dfrac{2\,M_{Z}^{2}\cos \alpha}{v}$\\
\hline
$Z\,Z\,h_2$ & $-\dfrac{2\,M_{Z}^{2}\sin \alpha}{v}$\\
\hline
$N_e\,N_{\mu}\,(N_{\tau})\,h_1$ & $\sqrt{2}
\sin \alpha\,h_{e\mu}\,(h_{e\tau})$\\
\hline
$N_e\,N_{\mu}\,(N_{\tau})\,h_2$ & $\sqrt{2}
\cos \alpha\, h_{e\mu} \,(h_{e\tau})$\\
\hline
$l\,\bar{l}\,h_1$ & $-\dfrac{M_{l}}{v} \cos \alpha$\\
\hline
$l\,\bar{l}\,h_2$ & $ \dfrac{M_{l}}{v} \sin \alpha$\\
\hline
$l\,\bar{l}\,Z_{\mu\tau}$ & ${\pm g_{\mu\tau}}\,
\gamma^\rho$($+$ for $\mu$, $-$ for $\tau$)\\
\hline
$\phi_{DM}\,\phi^{\dagger}_{DM}\,h_1$ &
$-(v\,\lambda_{Dh} \cos \alpha + \vmt
\lambda_{DH} \sin \alpha)$ \\
\hline
$\phi_{DM}\,\phi^{\dagger}_{DM}\,h_2$ &
$(v\,\lambda_{Dh} \sin \alpha - \vmt\,\lambda_{DH} \cos \alpha)$\\
\hline
$\phi_{DM}\,\phi^{\dagger}_{DM}\,Z_{\mu\tau}$ &
$n_{\mu\tau}\,g_{\mu\tau}\,(p_{2}-p_{1})^{\rho}$\\
\hline
$\phi_{DM}\,\phi^{\dagger}_{DM}\,h_1\,h_1$ & $-(\lambda_{Dh}\,
\cos^2 \alpha  + \lambda_{DH}\,\sin^2 \alpha)$\\
\hline
$\phi_{DM}\,\phi^{\dagger}_{DM}\,h_2\,h_2$ & $-(\lambda_{Dh}\,
\sin^2 \alpha  + \lambda_{DH}\,\cos^2 \alpha)$\\
\hline
$\phi_{DM}\,\phi^{\dagger}_{DM}\,h_1\,h_2$ &
$\sin \alpha \cos \alpha (\lambda_{Dh} - \lambda_{DH})$\\
\hline
$\phi_{DM}\,\phi^{\dagger}_{DM}\,Z_{\mu\tau}\,Z_{\mu\tau}$ & 
$2\,g^2_{\mu\tau} n^2_{\mu\tau}$\\
\hline
$\phi_{DM}\,\phi^{\dagger}_{DM} \phi_{DM}\,\phi^{\dagger}_{DM}$ &
$-4\,\lambda_{DM}$\\
\hline
\hline
\end{tabular}
\end{center}
\caption{All relevant vertex factors required for the computation
of DM annihilation as well as scattering cross sections.}
\label{tab3}
\end{table}     
\subsection{Results}
\label{sec-res}
We have computed the relic density of DM using micrOMEGAs
\cite{Belanger:2013oya} package and the implementation of the
present model in micrOMEGAS has been done using
the LanHEP \cite{Semenov:2008jy} package. For the
relic density calculation, we have considered the following benchmark 
values of the parameters related to the neutrino sector,
\begin{itemize}
\item Masses of the three heavy neutrinos: $M_{N_1} = 332.88$ GeV,
$M_{N_2} = 279.06$ GeV and $M_{N_3}=168.28\,{\rm GeV}$,
\item Yukawa couplings: $h_{e\mu}=2.44$ and $h_{e\tau}=1.28$.
\end{itemize} 
We have checked that these adopted values of right handed neutrino
masses and Yukawa couplings reproduce all the experimentally
measurable quantities of the neutrino sector within their 
$1\sigma$ range \cite{Capozzi:2016rtj}. 
Moreover like the previous section, here also we have used
our benchmark point $\mzmt=100$ MeV and $\gmt=9\times10^{-4}$,
which are required to explain the muon $(g-2)$ anomaly.
\begin{figure}[h!]
\centering
\includegraphics[angle=0,height=7.5cm,width=8.5cm]{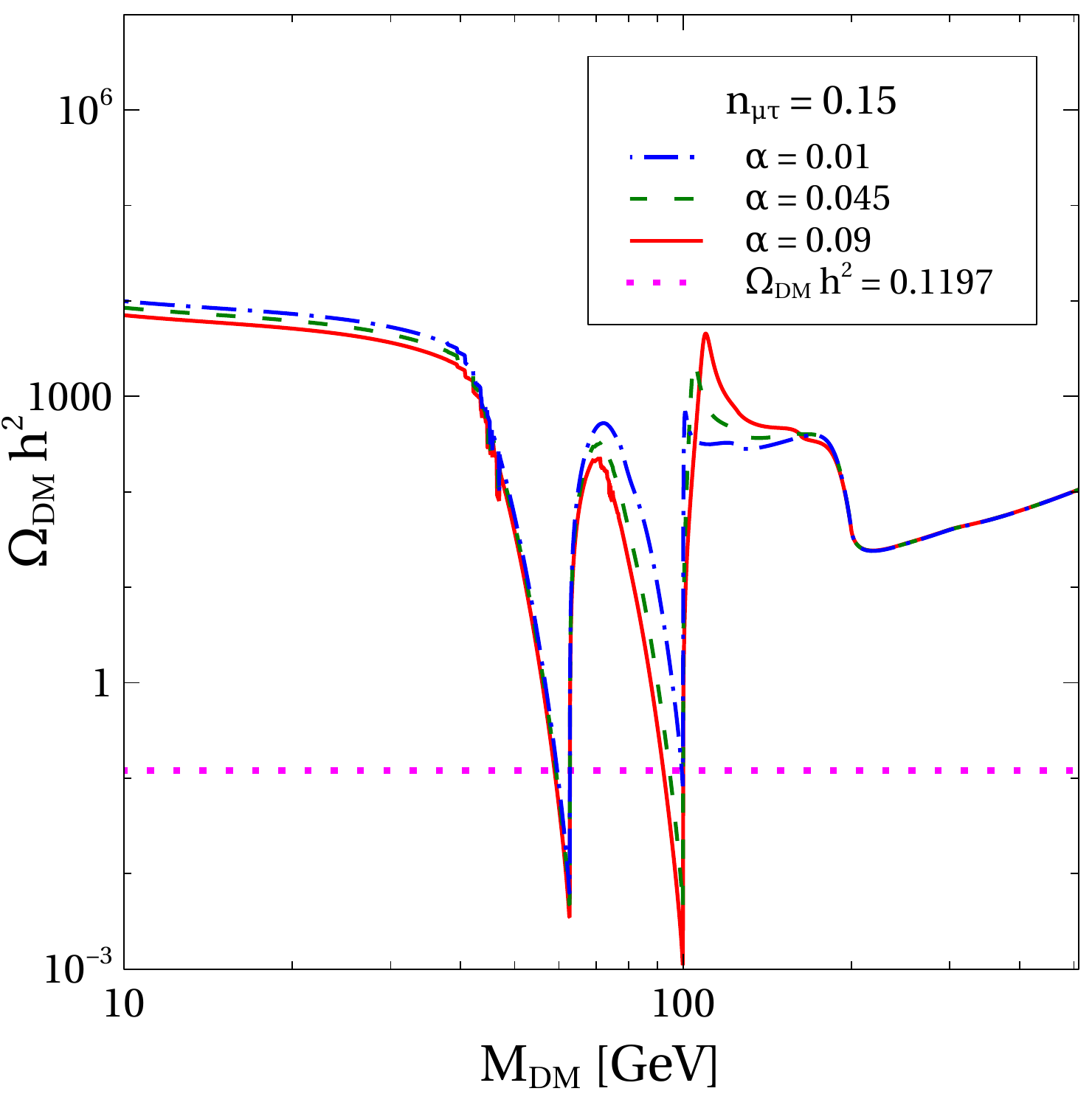}
\includegraphics[angle=0,height=7.5cm,width=8.5cm]{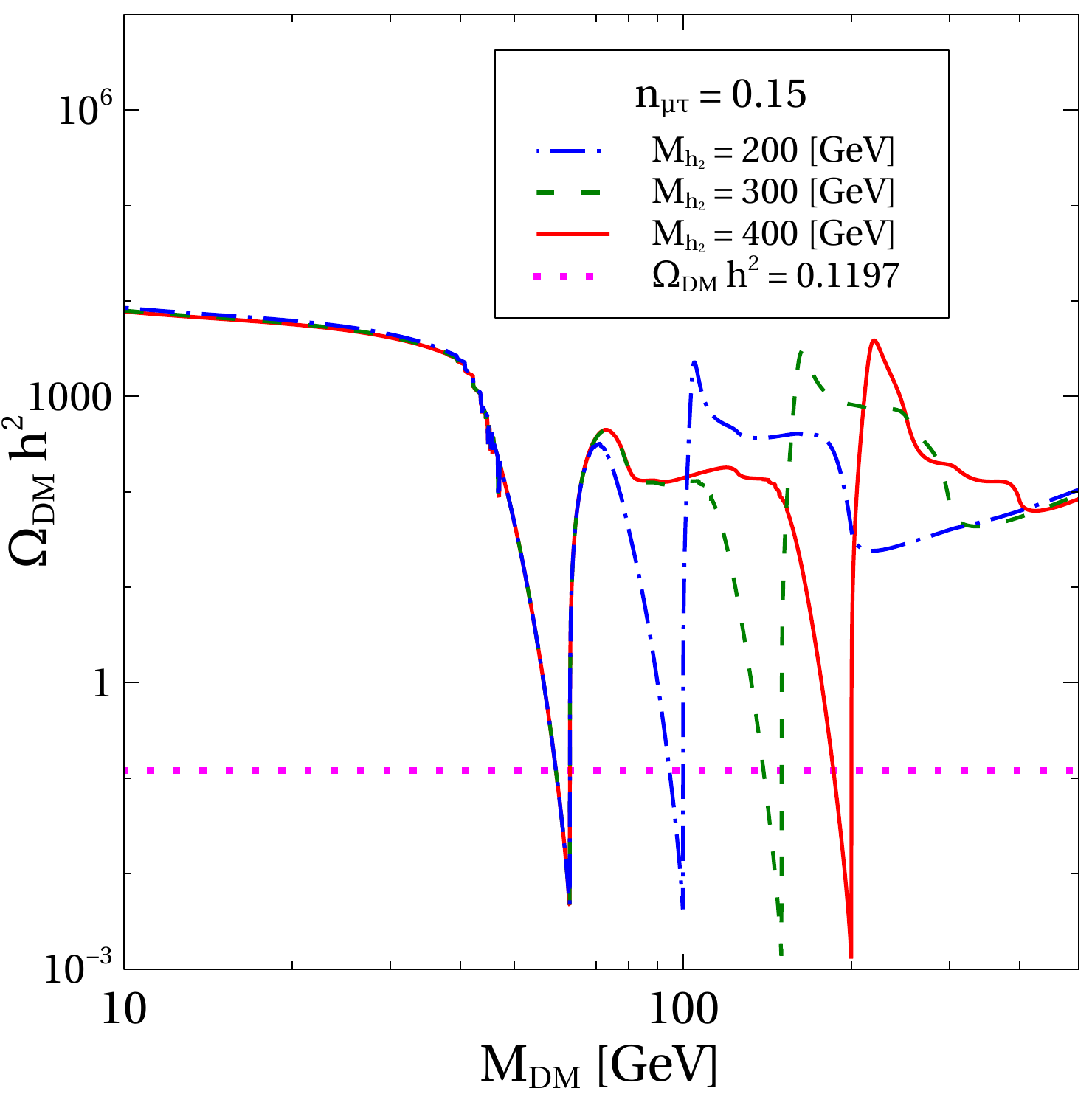}
\caption{Left (Right) Panel: Variation of relic density
$\Omega_{DM} h^{2}$ with respect to the DM mass $\mdm$ for
three different value of mixing angle $\alpha$ ($M_{h_2}$),
while other the values of parameters have been kept fixed at
$\ldH = 0.01$, $\ldh = 0.001$, and
$M_{h_2} = 200$ GeV ($\alpha = 0.045$ rad).}      
\label{dm1}
\end{figure}

In the left panel of Fig.\,\ref{dm1}, we show the variation of the 
DM relic density with its mass for three different values of the 
scalar mixing angle, $\alpha=0.01$ rad, $0.045$ rad and $0.09$ rad
\footnote{We have checked that these values of mixing
angle $\alpha$ are allowed by the LHC results on
Higgs signal strength \cite{Agashe:2014kda}
and invisible decay width \cite{Bechtle:2014ewa}.} respectively. 
From this plot it is clearly seen that DM relic density satisfies
the central value of Planck limit ($\Omega_{DM} h^2$ = 0.1197) only
around the two resonance regions where the mass of DM is nearly
equal to half of the mediator mass i.e. $\mdm \sim M_{h_i}/2$
($i=1$, 2). Therefore the first resonance occurs when
DM mass is around 62 GeV and it is due to the SM-like Higgs
boson $h_1$ while the second one is due to extra Higgs boson
$h_2$ of mass 200 GeV. 
Like the left panel of Fig.\,\ref{dm1}, the right panel also shows the variation of
$\Omega_{DM} h^2$ with $\mdm$ but in this case three different plots are generated
for three different values of $M_{h_2}=200$ GeV (blue dashed dot line),
$300$ GeV (green dashed line) and $400$ GeV (red solid line),
respectively. Similar to the left panel, here also the DM relic density
satisfies the Planck limit only around the resonance regions. However
in this plot, as we have varied the mass of $h_2$, therefore instead of getting
a single resonance region for $h_2$ (as in the left panel)
we have found three resonance regions at $\mdm\sim 100$ GeV, $150$ GeV
and 200 GeV for $M_{h_2} = 200$ GeV, 300 GeV and 400 GeV, respectively.
For all three cases the resonance due to
the SM-like Higgs boson $h_1$ occurs at the same
value of $\mdm\sim 62.5$ GeV as we have fixed the
mass of $h_1$ at 125.5 GeV. Plots in
both panels are generated for $\nmt=0.15$.   
\begin{figure}[h!]
\centering
\includegraphics[angle=0,height=7.5cm,width=8.5cm]{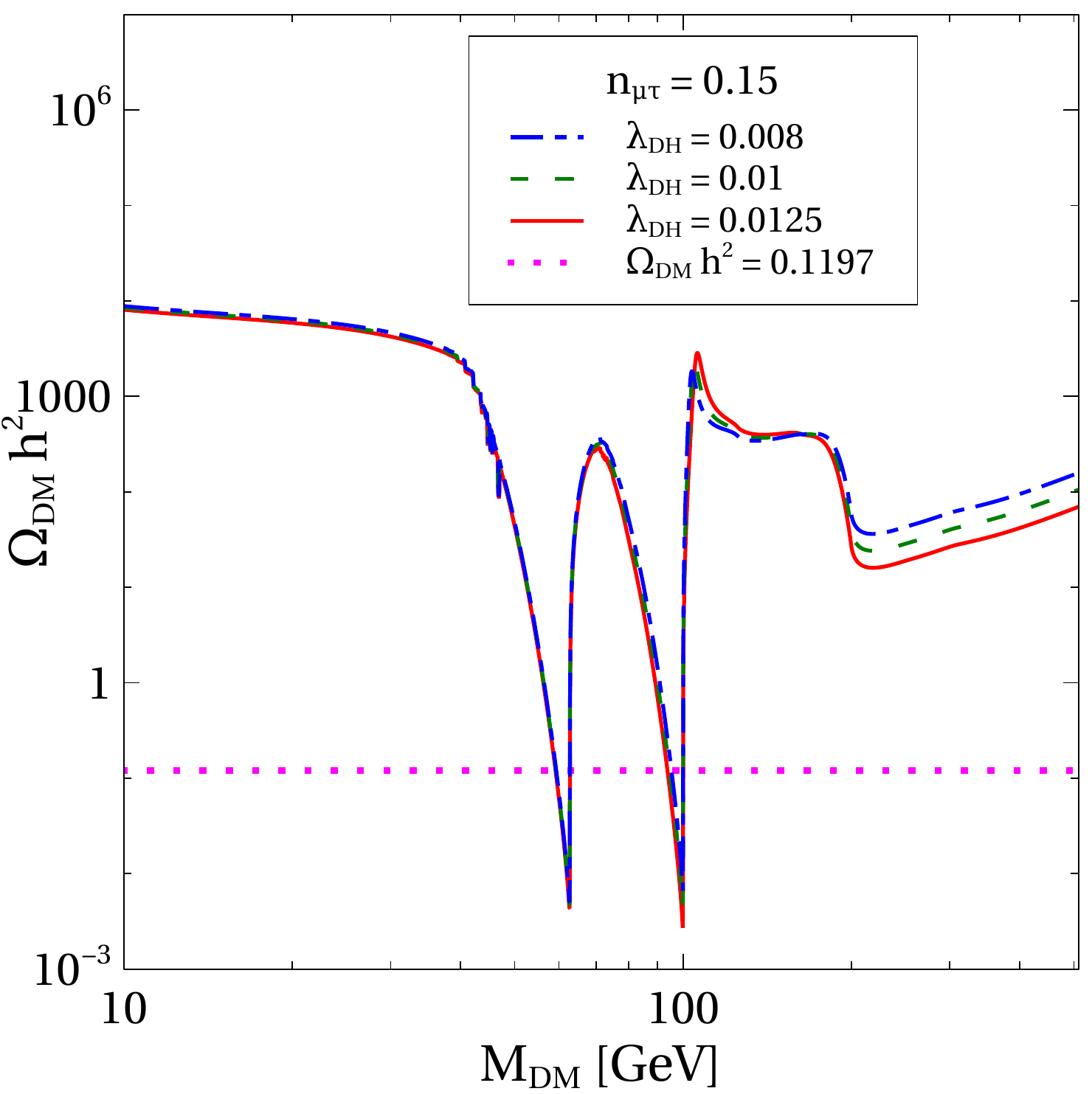}
\includegraphics[angle=0,height=7.5cm,width=8.5cm]{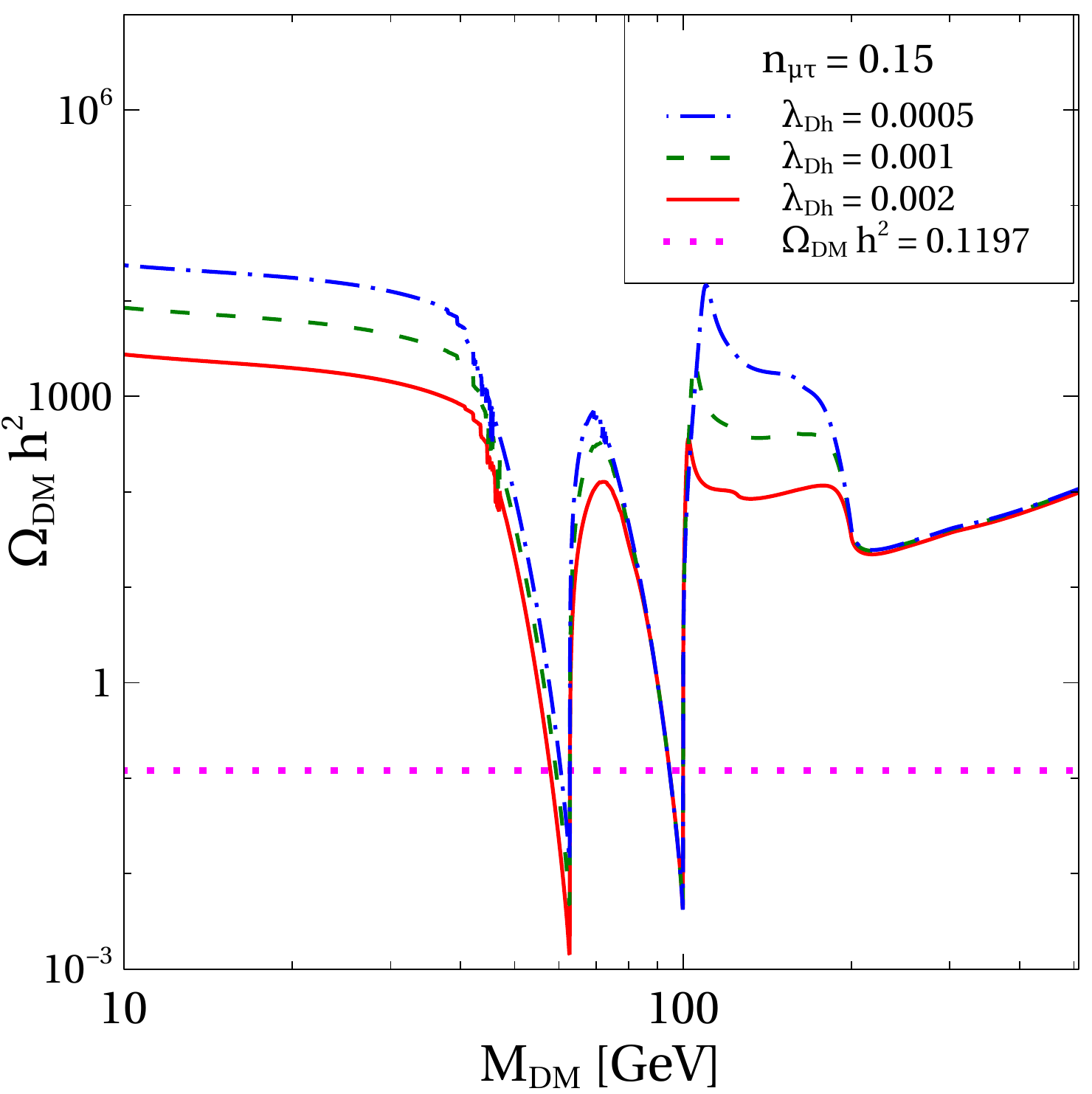}
\caption{Left (Right) Panel: Variation of relic density
$\Omega h^{2}$ with respect to the mass of the dark matter
$M_{DM}$ for three different value of $\lambda_{DH}$ ($\lambda_{Dh}$),
while other parameters value are kept fixed at $M_{h_2} = 200$ GeV,
$\alpha = 0.045$ rad and $\lambda_{Dh} = 0.001$ ($\lambda_{DH} = 0.01$).}
\label{dm2}
\end{figure}

Left and right panels of Fig.\,\ref{dm2} represent the variation
of relic density $\Omega_{DM} h^{2}$ with the dark matter mass $\dm$
for there different values of parameter $\lambda_{DH}$ and $\lambda_{Dh}$,
respectively. These plots also show the appearance of
two resonance regions due to the two mediating scalar bosons.
However, from this figure one can notice the effect of parameters
$\ldh$ and $\ldH$ on the DM relic density with respect to
the variation of $M_{DM}$. In the low mass region ($\mdm \la 80$ GeV),
SM-like Higgs boson mediated diagrams dominantly contribute to
the pair annihilation processes of $\dm$ and $\dmd$ while
the contribution of extra Higgs mediated diagrams become
superior for the high DM mass region ($\mdm\ga80$ GeV).
From the expression of $\dm\,\dmd\,h_1$ vertex
factor given in Table \ref{tab3}, one can see that the effect of the
parameter $\ldH$ on $\langle \sigma {\rm v} \rangle$ is mixing
angle suppressed (i.e. multiplied by $\sin {\alpha}$).
Therefore, in the left panel for low DM mass region
the effect of $\ldH$ to $\Omega_{DM} h^2$ is small.
On the other hand, in the expression of vertex factor
of $\dm\,\dmd\,h_1$, the parameter $\ldh$ appears with
$\cos \alpha$ and hence we see a considerable effect
of $\ldh$ on $\Omega_{DM} h^2$ in the right panel
(low DM mass region). For the extreme right
region of both panels ($\mdm\ga$ 200 GeV), the
dominant pair annihilation channel is $\dm\dmd\rightarrow h_2 h_2$.
Hence, the impact of $\ldH$ and $\ldh$ to $\Omega_{DM}h^2$
can well be understood from the expression of $\dm\dmd h_2h_2$
vertex factor (see Table \ref{tab3}). In the
intermediate region ($80\,{\rm GeV}<\mdm<200\,{\rm GeV}$),
$\dm \dmd \rightarrow W^+W^-$, $ZZ$ and $h_1h_1$
channels mainly contribute to DM relic density and in the
right panel for $100\,\text{GeV}<\mdm<200\,\text{GeV}$,
the variation of $\Omega_{DM} h^2$ with
respect to $\ldh$ resulting from DM pair annihilation
into $h_1h_1$ final state.  
\begin{figure}[h!]
\centering
\includegraphics[angle=0,height=7.5cm,width=8.5cm]{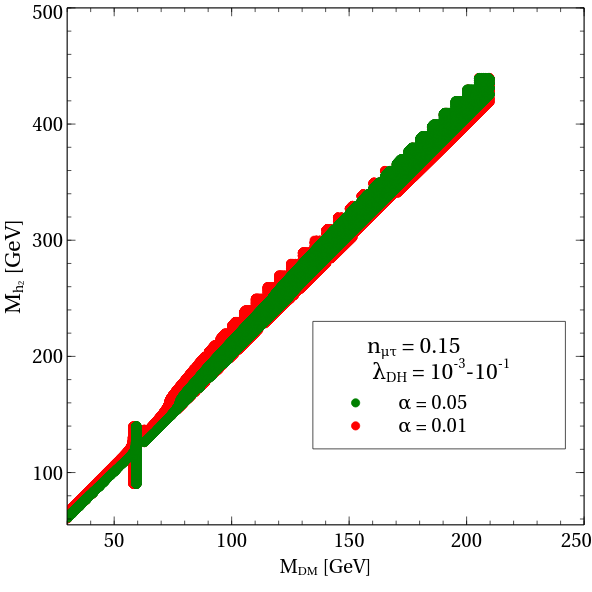}
\includegraphics[angle=0,height=7.5cm,width=8.5cm]{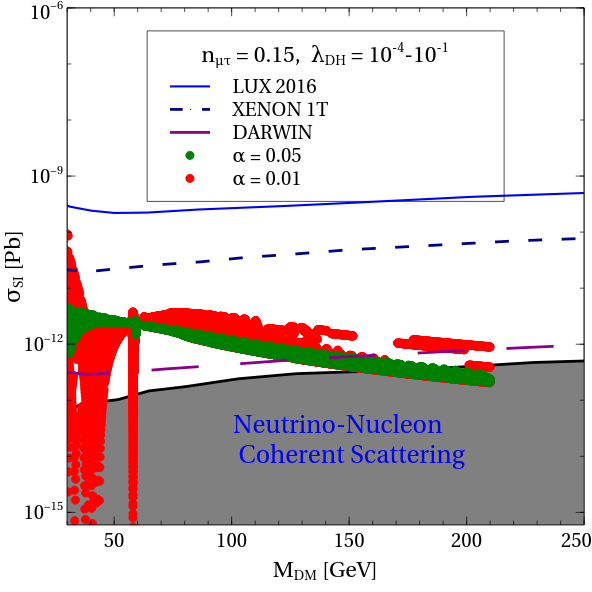}
\caption{Left Panel: Allowed values of $M_{h_2}$ with respect to
the variation of the dark matter mass $M_{DM}$ for two different
value of mixing angle $\alpha$. Right panel: Variation of spin
independent scattering cross sections of dark matter with its mass.
All the points in both plots satisfy the Planck limit on DM relic
density in $1\sigma$ range ($\Omega_{\rm DM} h^2= 0.1197\pm0.0022$
\cite{Ade:2015xua}) and these two plots are generated for
$\lambda_{Dh} = 0.001$.}
\label{dm3}
\end{figure}

In the left panel of Fig.\,\ref{dm3}, we show the allowed values of
$M_{h_2}$ which reproduce the correct DM relic density for the
variation of $\mdm$ in the range 30 GeV to 500 GeV. 
In this plot we have varied the mass of extra Higgs
boson $M_{h_2}$ in the range 60 GeV to 450 GeV and $\ldH$
from 0.001 to 0.1. From this plot it is evident that for
a particular value of dark matter mass the corresponding allowed
values of $M_{h_2}$ lie around $2\,\mdm$. The reason behind
this nature is that the relic abundance of dark matter
(both $\dm$ and $\dmd$) satisfies the observed DM density 
only around the resonance regions (when mediator mass
$M_{h_i} \sim 2\times\mdm$, $i=1,\,2$ see
Fig.\,\ref{dm1} and Fig.\,\ref{dm2}).
The allowed range of $M_{h_2}$
for a particular DM mass does not vary
much for the change of mixing angle
$\alpha$ from 0.01 rad (red coloured region)
to 0.05 rad (green colour region). Moreover,
we restrict $M_{h_2}$
upto 430 GeV to remain within the perturbative regime
($\lambda_{H}<4\,\pi$) and hence the 
relic density condition
is not satisfied beyond $\mdm=215\,{\rm GeV}$ 
Furthermore, near 
$M_{DM}\sim 60$ GeV, one can see that a broad range of $M_{h_2}$ values
are allowed, which indicates that in this region 
the SM-like Higgs contributes dominantly
giving the wide range of $M_{h_2}$ values for which the 
DM relic density is satisfied. Spin independent elastic
scattering cross section ($\sigma_{\rm SI}$) of DM
with with its mass has been plotted in the the
right panel of Fig.\,\ref{dm3} for two different
values of $\alpha =0.01$ rad (green coloured
region) and 0.05 rad (red coloured region) respectively.
This plot is also generated for $60\,{\rm GeV}\leq M_{h_2}\leq 430$ GeV,
$0.001 \leq \ldH \leq 0.1$ and $\ldh=0.001$
and all the points within the red and green coloured patch
satisfy the Planck result. For comparison with current
experimental limits on $\sigma^{\rm SI}$
from DM direct detection experiments
we have plotted the result of LUX-2016 (blue solid line)
in the same figure. Moreover, we have also shown the
predicted results from the ``ton-scale'' direct detection
experiments like XENON 1T \cite{Aprile:2015uzo} (blue dashed
line) and DARWIN \cite{Aalbers:2016jon} (long dashed purple line). 
From this figure it is evident that the validity of our 
model can be explored in near future by these ``ton-scale''
experiments.   
\subsection{Indirect detection: Fermi-LAT $\gamma$-ray excess from
the Galactic Centre}
Over the past few years, the existence of an unidentified excess of 
$\gamma$-rays with energy 1-3 GeV from the direction of the Galactic
Centre has been reported by several groups \cite{Goodenough:2009gk,
Hooper:2010mq, Boyarsky:2010dr, Hooper:2011ti, Abazajian:2012pn,
Hooper:2013rwa, Abazajian:2014fta, Daylan:2014rsa, Zhou:2014lva,
TheFermi-LAT:2015kwa, Agrawal:2014oha, Calore:2014nla} after analysing the 
Fermi-LAT publicly available data \cite{Atwood:2009ez}. There are some
astrophysical explanations such as unresolved point sources
(e.g. millisecond pulsar) around the GC which may be responsible for
this anomalous gamma-ray excess \cite{Lee:2015fea,
Bartels:2015aea}. However, the spectrum and
morphology of this gamma-ray excess is also very similar to that
expected from 
the annihilation \cite{Berlin:2014tja} or
decay (see \cite{Biswas:2015sva} and
references therein) of dark matter in the GC. In terms of an
annihilating DM scenario this excess can be well explained
by a dark matter of mass around $48.7^{+6.4}_{-5.2}$ GeV and 
with an annihilation cross section
$\langle {\sigma {\rm v}}_{b\bar{b}}\rangle
=1.75^{+0.28}_{-0.26}\times 10^{-26}$
cm$^3/$s into $b\bar{b}$ final state \cite{Calore:2014nla}. Thereafter
these $b$ quarks produce excess $\gamma$-ray from their
hadronization processes. The above quantities
$\mdm$ and $\langle {\sigma {\rm v}}_{b\bar{b}}\rangle$
depend on the specific choice of dark matter halo
profile. In Ref. \cite{Calore:2014nla} authors have used
an NFW halo profile \cite{Navarro:1996gj}
with index $\gamma=1.26$, $r_s=20$ kpc,
local dark matter density $\rho_{\odot}=0.4$ GeV/cm$^3$
and 
\begin{figure}[h!]
\includegraphics[height=15cm,width=10cm,angle=-90]{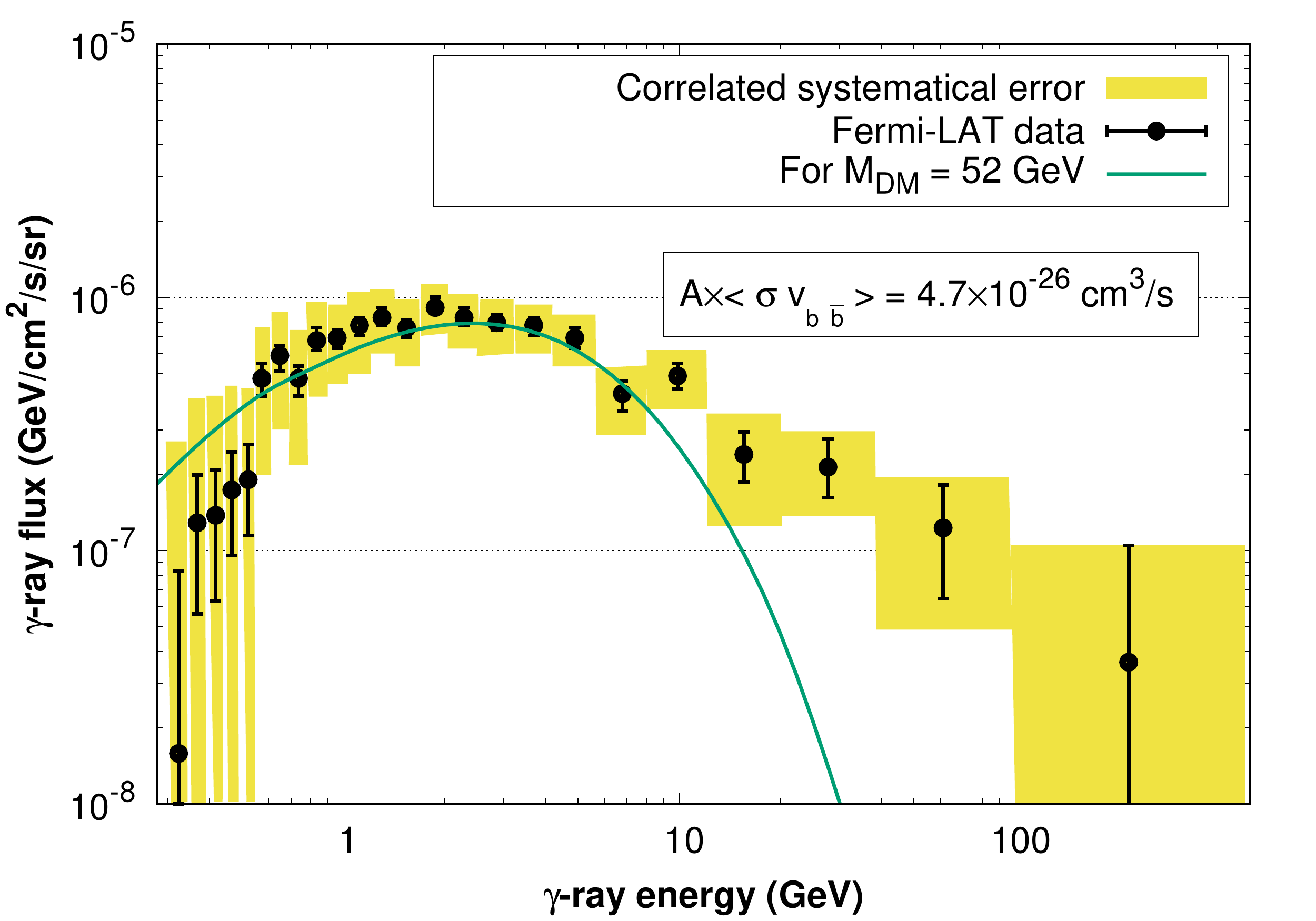}
\caption{Gamma-ray flux obtained from the pair annihilation
of $\dm$ and $\dmd$ at the Galactic Centre for $\mdm=52$ GeV,
$\langle {\sigma {\rm v}}_{b\bar{b}}\rangle=
3.856\times 10^{-26}\,{\rm cm}^3/{\rm s}$ and $\mathcal{A}=1.219$}
\label{gammaplot}
\end{figure}
a region of interest (ROI) around GC
where galactic latitude $b$, longitude $l$
vary in the range $2^0<|b|<20^0$, $|l|<20^0$
respectively during the analysis of Fermi-LAT data.
Since our knowledge about the exact
values of DM halo profile parameters such as
$\gamma$ and $\rho_{\odot}$ is limited, there
are some uncertainties in these profile parameters
and this can affect the calculated value of 
$\langle {\sigma {\rm v}}_{b\bar{b}}\rangle$. Due
to this uncertainty the allowed values of
annihilation cross section for the $b\bar{b}$ channel
can vary in the range $\mathcal{A} \times$
the best fit value of $\langle {\sigma {\rm v}}_{b\bar{b}}\rangle$
which is $1.75\times 10^{-26}\,{\rm cm}^3/{\rm s}$ while
$\mathcal{A}$ can be any number between 0.17 to 5.3 \cite{Calore:2014nla}.
For $\gamma=1.26$, $\rho_{\odot}=0.4\,{\rm GeV}/{\rm cm}^3$
and $r_s=20$ kpc and the value of $\mathcal{A}=1$, 
we have found that in the present $\lm-\lt$ symmetric
model with a DM candidate $\dm$ such explanation
of this anomalous gamma-excess
is indeed possible from the pair annihilation of
$\dm$ and $\dmd$ at the Galactic Centre. In our
earlier work \cite{Biswas:2016ewm} we have done a detailed computation
of $\gamma$-ray flux resulting from the annihilation of  
a complex scalar dark matter at the GC. Therefore, the
process of computing gamma-ray flux from the pair
annihilation of $\dm$ and $\dmd$ for the present
scenario is very similar to that work and hence 
these intermediated steps are not repeated here.
Note that since we are dealing with non-self-conjugate
dark matter, therefore, there will be an extra half factor in the expression
for the differential gamma-ray flux \cite{Cirelli:2010xx,
Biswas:2016ewm}. Hence in our case, the best fit value of
$\langle {\sigma {\rm v}}_{b\bar{b}}\rangle$
will be $3.50\times 10^{-26}\,{\rm cm}^3/{\rm s}$.
Following the same procedure given in \cite{Biswas:2016ewm}
we have found that, for the present model,
the excess gamma-rays flux observed by Fermi-LAT can be
reproduced for an annihilating dark matter of
mass $\mdm=52$ GeV and $\langle {\sigma {\rm v}}_{b\bar{b}}\rangle
= 3.856 \times 10^{-26}\,{\rm cm^3}/{\rm s}$. In this
case, DM annihilation to $b\bar{b}$ channel dominantly
occurs through the resonance of extra Higgs boson ($h_2$)
with resonating mass $M_{h_2}=104.025$ GeV and
coupling parameters $\ldH=0.01$, $\ldh=0.001$
and scalar mixing angle $\alpha=0.045$ rad.

In Fig.\,\ref{gammaplot}, green solid line represents
the $\gamma$-ray flux that we have computed
for a $\mdm=52$ GeV while the value of $b\bar{b}$ annihilation cross
section is $3.856 \times 10^{-26}\,{\rm cm^3}/{\rm s}$. The
correlated systematic errors are represented by
the yellow boxes while the Fermi-LAT uncorrelated
statistical uncertainties are shown by the black error bars 
taken from \cite{Calore:2014xka}.
We have found that in order to reproduced the Fermi-LAT observed
$\gamma$-ray flux for a 52 GeV non-self-conjugate DM,
the quantity $\mathcal{A}\times\langle {\sigma {\rm v}}_{b\bar{b}}\rangle$
must be $4.7 \times 10^{-26}\,{\rm cm^3}/{\rm s}$ \cite{Biswas:2016ewm}.
This requires DM halo profile error parameter $\mathcal{A}$
to be $\sim 1.22$, well inside its allowed range
between 0.17 to 5.3 \cite{Calore:2014nla}. 
\section{Summary and Conclusion}
\label{sandc}
Although 
Standard Model (SM) is a well established theory of elementary particle physics, it 
cannot explain 
the muon (${g-2}$) anomaly,
the small neutrino masses and peculiar mixing pattern, 
and the existence of Dark Matter (DM). Therefore, the SM has to be extended to 
explain these observational evidences. 
In the present work
we have extended the SM gauge group SU(3)$\times$SU(2)$_{L} \times$U(1)$_{Y}$ by a
local $U(1)_{L_{\mu}-L_{\tau}}$ gauge group. Since we require $U(1)_{L_{\mu}-L_{\tau}}$ 
to be local, we get an extra gauge boson, $Z_{\mu\tau}$. 
One of the most appealing aspects of 
the gauged $U(1)_{L_{\mu}-L_{\tau}}$ extension of the SM is that 
it does not introduce any anomaly in the theory \cite{He:1990pn, He:1991qd, Ma:2001md}. 
We introduce a scalar with non-trivial 
$\lm-\lt$ number which picks up a VEV, breaking the $U(1)_{L_{\mu}-L_{\tau}}$ symmetry 
spontaneously and making $Z_{\mu\tau}$ massive. This extra massive $Z_{\mu\tau}$ 
provides additional contributions to the magnetic moment of the muon, which 
can explain the observed data on muon (${g-2}$) for $\zmt$ of 
$\mathcal{O}$ (100 MeV) and low
values of gauge coupling $\gmt \la 10^{-3}$. 
We fixed the value
of $g_{\mu\tau}$ and $M_{Z_{\mu\tau}}$ such that they are 
allowed by the neutrino trident
process \cite{Altmannshofer:2014pba} and calculated the 
muon (${g-2}$) to within $3.2\,\sigma$ of the measured value. We kept 
$g_{\mu\tau}$ and $M_{Z_{\mu\tau}}$ fixed at these values throughout the rest 
of the paper. 

The $\lm-\lt$ symmetry, being also a flavor symmetry, provides a natural 
way of explaining the peculiar mixing pattern of the light neutrinos. 
We added to the particle content, three right-handed neutrinos ($N_{e}$, $N_{\mu}$, $N_{\tau}$)
and generated small neutrino masses naturally through 
the canonical Type-I seesaw mechanism. The $N_{e}$, $N_{\mu}$, $N_{\tau}$ 
are given $\lm-\lt$ flavor numbers, making the right-handed neutrino mass matrix 
and as a result the light Majorana neutrino mass matrix $\mu-\tau$ symmetric. 
This leads to $\theta_{23}=\pi/4$ and $\theta_{13}=0$, inconsistent with the 
neutrino oscillation data. However, when the $\lm-\lt$ symmetry gets spontaneously broken, 
it generates additional terms in the right-handed and consequently light neutrino mass 
matrix giving a good explanation of the global neutrino oscillation data. We scanned 
the five-dimensional model parameter space of our model and found the regions 
of this space that are consistent with the allowed neutrino oscillation parameters within 
their $3\sigma$ ranges. We discussed the correlations between the model parameters. 
We also presented the oscillation parameters predicted by our model. In particular, we 
showed that our model can explain the observed value of $\theta_{13}$ very naturally, 
predicts a value of 
$\theta_{23}$ that is not maximal, does not distinguish between the two octants of 
$\theta_{23}$ and predicts the Dirac $\delta_{CP}$ phase to be very close to 0. 
Hence our model predicts that no discernible CP violation will be observed 
in the long baseline experiments.

We next introduced another complex scalar $\phi_{DM}$ which 
does not take a VEV and hence is a good candidate for DM. The stability of 
this complex scalar is ensured by giving it a suitable $\lm-\lt$ charge, making it impossible to 
write any decay terms in the Lagrangian, even after the $\lm-\lt$ symmetry is 
broken spontaneously. We showed that due to the very small gauge coupling 
$\gmt$ required to explain the anomalous muon (${g-2}$) data, the $\zmt$-portal 
diagrams do not contribute to the DM phenomenology. The relic abundance and 
signature of our model in direct and indirect experiments come through the Higgs portal. 
We calculated the relic abundance of DM in this model and 
showed that the observational constraints from Plank can be satisfied for the two
resonance regions corresponding to the scenario where $M_{DM}\simeq M_{h_1}/2$ and 
$M_{DM}\simeq M_{h_2}/2$, respectively, where $M_{h_1}$ and $M_{h_2}$ are 
the masses of ${h_1}$ 
and ${h_2}$, the two Higgs scalars in our model. We presented the prediction 
of our model in forthcoming direct detection experiments and showed that 
for a wide range of model parameter space, 
XENON 1T and DARWIN could see a positive signal for $\phi_{DM}$. Likewise, they 
can constrain large parts of the model parameter in case they do not observe any 
WIMP signal. We also showed that for $\phi_{DM} \simeq 52$ GeV, our model 
can explain the galactic centre gamma ray excess in the $1-3$ GeV range 
observed by FermiLAT. 

In conclusion, we propose a gauged $\lm-\lt$ extension of the SM with two 
additional scalars and three additional right-handed neutrinos. This model 
can explain the anomalous muon $({g-2})$ data, small neutrino masses 
and peculiar mixing pattern, and provides a viable dark matter candidate. 
It can explain the relic abundance as well as the galactic centre gamma ray excess 
while satisfying all other experimental bounds. It also predict no CP violation in 
neutrino oscillation experiments. This model is phenomenologically 
rich and predictive and should be testable in forthcoming high energy physics 
experiments, including collider experiments, dark matter experiments 
as well as neutrino oscillation experiments.
 
\section{Acknowledgements} One of the authors A.B. wants to
thank Mainak Chakraborty for some valuable discussions.
A.B. also likes to thank Arindam Mazumdar for his help
in gnuplot. The authors would like to thank the Department
of Atomic Energy
(DAE) Neutrino Project under the XII plan of Harish-Chandra
Research Institute. SK and AB also acknowledge the cluster
computing facility at HRI (http://cluster.hri.res.in).
This project has received funding from the
European Union’s Horizon 2020 research and
innovation programme InvisiblesPlus RISE under the
Marie Skłodowska-Curie grant agreement No 690575.
This project has received funding from the European
Union’s Horizon 2020 research and innovation
programme Elusives ITN under the Marie Sklodowska-Curie
grant agreement No 674896.


\end{document}